\documentclass[11pt]{article}
\usepackage{jheppub}

\usepackage{amsmath}
\usepackage{graphicx}
\usepackage{epstopdf}
\usepackage{forest}	
\usepackage{subfigure}
\usepackage{multirow}
\usepackage{caption}
\usepackage{amsfonts,amssymb}
\usepackage[numbers,sort&compress]{natbib}
\usepackage{makecell}

\allowdisplaybreaks[4]
\graphicspath{{./figs/}}

\preprint{USTC-ICTS/PCFT-22-23}
\title{Gluonic evanescent operators: two-loop anomalous dimensions}
\author[a]{Qingjun Jin,}
\emailAdd{qjin@gscaep.ac.cn}
\author[b,c]{Ke Ren,}
\emailAdd{renk9@mail.sysu.edu.cn}
\author[c,d,e,f]{Gang Yang,}
\emailAdd{yangg@itp.ac.cn}
\author[g,c]{and Rui Yu}
\emailAdd{yurui@csrc.ac.cn}
\affiliation[a]{Graduate School of China Academy of Engineering Physics, No.~10 Xibeiwang East Road, Haidian District, Beijing, 100193, China}
\affiliation[b]{School of Physics and Astronomy, Sun Yat-Sen University, Zhuhai 519082, China}
\affiliation[c]{CAS Key Laboratory of Theoretical Physics, Institute of Theoretical Physics, \\Chinese Academy of Sciences, Beijing 100190, China}
\affiliation[d]{School of Fundamental Physics and Mathematical Sciences, Hangzhou Institute for Advanced Study, UCAS, Hangzhou 310024, China}
\affiliation[e]{International Centre for Theoretical Physics Asia-Pacific, Beijing/Hangzhou, China}
\affiliation[f]{Peng Huanwu Center for Fundamental Theory, Hefei, Anhui 230026, China}
\affiliation[g]{Beijing Computational Science Research Center, Beijing 100193, China}

\abstract{
Evanescent operators are a special class of operators that vanish in four-dimensional spacetime but are non-zero in $d=4-2\epsilon$ dimensions.
In this paper, we continue our systematic study of the evanescent operators in the pure Yang-Mills theory and focus on their two-loop renormalization.
We develop an efficient strategy to compute the two-loop divergences of form factors of high-dimensional and high-length operators by combining the $d$-dimensional unitarity method and the improved tensor reduction techniques.
Two-loop anomalous dimensions are obtained for the dimension-10 basis in the planar YM theory, 
for which both the $\overline{\text{MS}}$ scheme and the finite-renormalization scheme are used.
We verify that the two-loop anomalous dimensions are the same in these two schemes at the Wilson-Fisher conformal fixed point.
Our computation shows that the evanescent operators are indispensable in order to obtain the correct two-loop anomalous dimensions. 
This work provides a first computation of the two-loop anomalous dimensions of the complete set of dimension-10 operators. 
The method we use is also expected to provide an efficient strategy for the two-loop renormalization of general high-dimensional operators.
}

\begin{document}
\maketitle

\section{Introduction}
Dimensional regularization (DR) \cite{tHooft:1972tcz} is a widely used regularization scheme to regularize divergences in the loop Feynman integrals, where the spacetime dimension is continued to non-integer $d=4-2\epsilon$ dimensions. In general $d$ spacetime dimensions, there exists a special class of so-called evanescent operators that vanish in four dimensions (\emph{i.e.}~in the limit $\epsilon\to 0$). 
In our previous work \cite{Jin:2022ivc}, we discussed the systematic construction of the gluonic evanescent operators and performed the one-loop renormalization. Based on those results, we continue to study the two-loop renormalization of evanescent operators in this paper. 

Starting from the two-loop order, evanescent operators are expected to have non-negligible physical effects.
This has been studied a long time ago for the four-fermion type evanescent operators
in four-dimensional \cite{Buras:1989xd,Dugan:1990df,Herrlich:1994kh,Buras:1998raa} and two-dimensional \cite{Bondi:1989nq,Vasiliev:1997sk,Gracey:2016mio} theories. There is an important difference between the gluonic evanescent operators and the four-fermion type evanescent operators. In the four-fermion case, there are infinite number of operators with the same canonical dimensions which in general can mix with each other, and thus a certain special choice of the operators \cite{Buras:1989xd,Dugan:1990df,Herrlich:1994kh} or some truncation of the basis \cite{DiPietro:2017vsp} are required.
In comparison, the gluonic operator basis at a given mass dimension contains only a finite number of operators,
and this allows us to study the evanescent effects in a concrete setup.
For example, we can compute the anomalous dimensions with a general choice of a basis and also address the possible issue of the scheme dependence without any ambiguity.
We note that the finiteness of the basis number also happens for evanescent operators in the scalar theories \cite{Hogervorst:2015akt,Cao:2021cdt}.
Evanescent effects in gravitational theories were also considered in \cite{Bern:2015xsa, Bern:2017puu}.

In this paper, we obtain the two-loop anomalous dimensions for the dimension-10 operators (operators of classical dimensions 10) in the planar YM theory.\footnote{Since the anomalous dimensions of dimension-10 operators itself includes the results of lower-dimension operators, our results also provide for the first time the anomalous dimensions of the dimension-8 operators.} 
There are 36 independent operators and six of them are evanescent.
We use two different schemes to perform the renormalization: the modified minimal subtraction ($\overline{\text{MS}}$) scheme \cite{Bardeen:1978yd} and the finite renormalization scheme \cite{Buras:1989xd,Bondi:1989nq}. The anomalous dimensions are scheme dependent because of the effect of the non-zero beta function. To further check our results, we consider the Wilson-Fisher (WF) conformal fixed point \cite{Wilson:1971dc} of the pure YM theory where the anomalous dimensions should be scheme independent. We find that the two schemes indeed give the
same results, which provides a non-trivial consistency check of our computation.

We would like to stress that the two-loop computation for the high-dimensional and
high-length operators in the YM theory is also a
challenging technical problem. For example, to calculate the anomalous dimensions of operators
at classical dimension 10, two-loop four-point and five-point form factors are required. We employ a strategy that combines the unitarity method \cite{Bern:1994cg,Bern:1994zx,Britto:2004nc} and the integration by parts (IBP) reduction \cite{Chetyrkin:1981qh,Tkachov:1981wb}. Importantly, since we aim to study the evanescent operators, the computation must be performed in $d$ dimensions. Therefore, it is natural to use the $d$-dimensional unitarity cut method and work in the conventional dimensional regularization (CDR). 
A technical challenge in computing the form factors of high-length operators is to perform non-trivial tensor integral reduction. We find that this reduction can be done relatively efficiently by a modified Passarino-Veltman reduction
method \cite{Passarino:1978jh,Kreimer:1991wj}.
The efficiency of the program can be further enhanced by using the numerical reconstruction method.
Since the operator basis is finite, it is straightforward to reconstruct the analytic renormalization matrix using a finite set of numerical data. In this way, we obtain the analytic two-loop anomalous dimensions. 
The strategy we employ is expected to also provide an efficient framework for
the two-loop renormalization of general high-length operators.

The paper is arranged as follows. In Section~\ref{sec:prepare}, we introduce necessary notations and the dimension-10 operator basis. In Section~\ref{sec:renorm}, we describe the renormalization procedure and give the explicit renormalization formulas in the $\overline{\text{MS}}$ scheme and the finite renormalization scheme. In Section~\ref{calcbare}, we explain our calculation of the loop form factors using the $d$-dimensional unitarity method, and a detailed description of how to deal with the challenging tensor reduction is also given.  In Section~\ref{sec:result}, we present our results, including the renormalization matrices and the anomalous dimensions. A summary and discussion are given in Section~\ref{sec:discuss}. The explicit
basis of the dimension-10 operators is given in Appendix \ref{all dim-10}. The two-loop mixing between physical operators is presented in Appendix~\ref{zppresult}.

\section{Preparation}\label{sec:prepare}
In this section, we first set up basic notations and our conventions for local operators in the pure YM theory in Section~\ref{ymnotation}. Then in Section~\ref{allopers}, we define the physical and evanescent operators, and we also present the dimension-10 operator basis .

\subsection{Setup in the YM theory}\label{ymnotation}
The Lagrangian of the pure YM theory is\footnote{For simplicity, in this paper we will not distinguish upper and lower Lorentz indices. For example, $\eta^{\mu\nu}$ and $\delta^{\mu}_\nu$ are regarded as equivalent. This will not cause any problem in the flat spacetime.}
\begin{align}
\mathcal{L}=-\frac{1}{2}\text{tr}(F_{\mu\nu}F_{\mu\nu}) = -\frac{1}{4} F^a_{\mu\nu}F^a_{\mu\nu}\,,
\end{align}
where $F_{\mu\nu}\equiv F^a_{\mu\nu} T^a$, and $T^a,\ \text{with } a = 1,..,N_c^2-1$, is the SU($N_c$) generator. The field strength tensor $F^a_{\mu\nu}$ is defined as
\begin{align}
F^a_{\mu\nu}=\partial_\mu A_\nu^a-\partial_\nu A_\mu^a+g_0 f^{abc}A_\mu^bA_\nu^c\,,
\end{align}
where $A_\mu^a$ is the gauge field, $f^{abc}$ is the structure constant and $g_0$ is the bare coupling. The Einstein summation is assumed for the repeated indices.

The covariant derivative is defined to be
\begin{align}
D_\mu=\partial_\mu-\text{i}g_0A_\mu^aT^a\,.
\end{align}
The action of the covariant derivative is
\begin{align}
D_\mu X\equiv [D_\mu,X]\,,
\label{Dact}
\end{align}
given that $X=X^aT^a$. The commutator of two $D_\mu$'s reads
\begin{align}
[D_\mu,D_\nu]=-\text{i}g_0F_{\mu\nu}\,.
\label{Dcommute}
\end{align}
The equation of motion reads
\begin{align}
D_\mu F_{\mu\nu}=0\,.
\label{eom}
\end{align}
Another important relationship is the Bianchi identity:
\begin{align}
D_{\mu}F_{\rho\sigma}=D_{\sigma}F_{\rho\mu}+D_{\rho}F_{\mu\sigma}\,.
\label{bianchi}
\end{align}

We aim to study the gauge-invariant local scalar operators. For simplicity, in this paper we will take the large $N_c$ limit. Consequently, we only need to consider the single-trace operators. 
We will often abbreviate the Lorentz indices by integer numbers, such as
\begin{align}
\text{tr}(F_{12}F_{12})\equiv \text{tr}(F_{\mu_1\mu_2}F_{\mu_1\mu_2})\,.
\end{align}
It is conventional to classify the operators according to their classical dimensions. For example, we say that $\text{tr}(F_{12}F_{12})$ is a dimension-4 operator.

Given an operator {$O$}, its $n$-point form factor is defined to be
\begin{align}
\mathcal{F}_{O,\text{$n$ gluons}}=\int \text{d}^dx e^{-{\text{i}}q\cdot x}\langle \text{g}_1\cdots,\text{g}_n |O|0\rangle\,,
\end{align}
where $\text{g}_i$ denotes an external gluon carrying an on-shell momenta $p_i$, and $q=\sum_i {p_i}$ is the off-shell momentum associated to the operator. See \emph{e.g.}~\cite{Yang:2019vag} for a recent introduction of form factors. All the gluons are assumed to be outgoing in this paper. 
For an operator $O$, there exists an integer number $m$ such that {its tree-level $n$-point form factors vanish unless $n\geq m$.}
We call the $m$-gluon form factor the minimal form factor of this operator. Similarly, an $(m+1)$-point form factor is called a next-to-minimal form factor, and a $(m-1)$-point form factor is called a sub-minimal form factor. The tree-level color-ordered minimal form factor can be achieved straightforwardly by the dictionary (see \emph{e.g.}~\cite{Jin:2022ivc})
\begin{align}
F^{\mu\nu}\to \text{i}(p^\mu e^\nu-p^\nu e^\mu),\qquad D^{\mu}\to \text{i}p^\mu\,,
\end{align}
where $e^\mu$ is the polarization  of a gluon and $p^\mu$ is the momentum of a gluon. By definition, there is no tree-level sub-minimal form factor. For the tree-level next-to-minimal form factors, compact formulas are given in \cite{Jin:2022ivc}.

The length of an operator is defined to be the number of gluons of its minimal form factor.\footnote{In most cases, one can simply take the length as equal to the number of $F_{\mu\nu}$ in an operator. However, this may cause ambiguity sometimes. For example, the operator $\text{tr}([D_1,D_2]F_{34} F_{41} F_{23})=-ig_0 \text{tr}([F_{12},F_{34}] F_{41} F_{23})$ do not have a definite length by counting the number of $F_{\mu\nu}$; while using the above definition based on the minimal form factor, this operator is of length 4.} For example, the operator
\begin{align}
\text{tr}(D_3D_4F_{12}F_{23}F_{41})+D^4\big(\text{tr}(F_{12}F_{12})\big)\,,
\label{opexample}
\end{align}
contains a length-3 and a length-2 monomial operators.
The length of the whole operator is two, since the minimal tree form factor is a two-point form factor. 
In this work, we take the convention such that each length-$L$ monomial operator is implicitly multiplied by a factor $(-\text{i}g_0)^{L-2}$. For example, \eqref{opexample} means $(-\text{i}g_0) \text{tr}(D_3D_4F_{12}F_{23}F_{41})+D^4\big(\text{tr}(F_{12}F_{12})\big)$. This convention is due to the fact that a strength tensor can be interpreted as a commutator of two $D_{\mu}$'s which involves a factor $-\text{i}g_0$, and the shortest operator in this work is of length two. Under charge conjugation, the color trace of a monomial operator is reversed, together with a factor $(-1)^L$ with $L$ the length of the monomial operator \cite{Bardeen:1969md}. For example,
\begin{align}
{\rm tr}(D_3D_4F_{12}F_{23}F_{41}) \to - {\rm tr}(F_{41}F_{23}D_3D_4F_{12})\,.
\end{align}
An eigenstate operator under change conjugation with $+\ (-)$ C-parity is said to be C-even (C-odd).

An operator before renormalization is called a bare operator, denoted as $O_\text{b}$. Since we aim to study the evanescent operators, we will adopt the CDR scheme to regularize the divergences of a bare operator, with loop momenta and external momenta all in $d$ spacetime dimensions. Given a set of bare operators $\{O_{\text{b},j}\}$, a renormalized operator $O_i$ can be written as 
\begin{align}
{O}_i=Z_i^{~j}{O}_{\text{b},j}\,.
\label{renormoperator}
\end{align}
A matrix element $Z^{~j}_i$ represents an operator mixing from $O_{i,\text{b}}$ to $O_{j,\text{b}}$. In general, operators of the same classical dimension can mix with each other. The dilatation matrix $\mathcal{D}$ is defined as
\begin{align}
\mathcal{D}\equiv -\mu\frac{\text{d}Z}{\text{d}\mu}Z^{-1}=\sum_{l=1}\left(\frac{\alpha_s}{4\pi}\right)^l\mathcal{D}^{(l)}\,.
\label{overallgamma}
\end{align}
The eigenvalues of the dilatation matrix are the anomalous dimensions, denoted as $\gamma$.

\subsection{Physical and evanescent operators}\label{allopers}
We call an operator a physical operator if it does not vanish in four spacetime dimensions. Otherwise, we call it an evanescent operator. For example, the operator
\begin{align}
\delta^{12345}_{6789(10)}\text{tr}(F_{12}F_{34}F_{56}F_{78}F_{9(10)})\,,
\label{expforeva}
\end{align}
is an evanescent operator, where the generalized Kronecker symbol is defined to be
\begin{equation}
\delta^{\mu_1..\mu_n}_{\nu_1...\nu_n}= {\rm det}(\delta^\mu_\nu) =
\left|
\begin{matrix}
\delta^{\mu_1}_{\nu_1} & \ldots  & \delta^{\mu_1}_{\nu_n} \\
\vdots &  & \vdots\\
\delta^{\mu_n}_{\nu_1} & \ldots & \delta^{\mu_n}_{\nu_n}
\end{matrix}
\right|\,.
\end{equation}
One can see that the rank-5 Kronecker symbol guarantees the vanishing of \eqref{expforeva} in four spacetime dimensions. More details of systematic construction of evanescent operators can be found in \cite{Jin:2022ivc}.

We will choose the dimension-10 operators as concrete examples to study the effects of evanescent operators on the two-loop renormalization. 
As mentioned above, we will consider the large $N_c$ limit and thus we focus only on the single-trace operators. The arrangement of our operator basis is summarized in Table~\ref{opertable}. The basis is first divided into two classes: the physical ones and the evanescent ones. Within each class, we classify the operators according to their C-parities and then into different lengths. The physical operators can be further arranged into different helicity sectors---within each helicity sector, the operators' tree minimal form factors are only non-vanishing in the corresponding helicity configuration (and the conjugate configuration). For example, $\text{tr}(F_{12}F_{12})$ is in the $(-)^2$ sector and its tree-level minimal form factor is non-vanishing only if the helicities are $(-)^2$ or $(+)^2$.  
The explicit definitions of the operators are given in Appendix~\ref{all dim-10}.
A similar basis has been given in \cite{Jin:2022ivc}.
Here an improvement is made: the basis is reorganized in a form such that the total derivative operators are separated explicitly, see Appendix~\ref{all dim-10} for further details. This reorganization will help to analyze the comparison of the anomalous dimensions in different renormalization schemes in Section~\ref{getZ}.

\begin{table}
\centering
	\begin{tabular}{|c|l|l|l}
		\hline
		\multirow{6}{*}[-40pt]{30 physical}                        & \multirow{4}{*}[-20pt]{24 C-even} & 1 length-2  &  \multicolumn{1}{l|}{$O_1:$ $(-)^2$}                                                                                                  \\ \cline{3-4}
		&                            & 4 length-3  & \multicolumn{1}{l|}{\begin{tabular}[c]{@{}l@{}}$O_2,O_3:$ $(-)^3$\\ $O_4,O_5:$ $(-)^2+$\end{tabular}}                                                \\ \cline{3-4}
		&                            & 15 length-4 & \multicolumn{1}{l|}{\begin{tabular}[c]{@{}l@{}}$O_6$--$O_{10}:$ $(-)^4$\\$O_{11}$--$O_{14}:$ $(-)^3+$\\ $O_{15}$--$O_{20}:$ $(-)^2(+)^2$\end{tabular}}  \\ \cline{3-4}
		&                            & 4 length-5  & \multicolumn{1}{l|}{\begin{tabular}[c]{@{}l@{}}$O_{21},O_{22}:$ $(-)^5$\\$O_{23},O_{24}:$ $(-)^3(+)^2$\end{tabular}}         \\ \cline{2-4}
		& \multirow{2}{*}[-13pt]{6 C-odd}   & 1 length-3  & \multicolumn{1}{l|}{$O_{25}:$ $(-)^2+$}                                                                                                \\ \cline{3-4}
		&                            & 5 length-4  & \multicolumn{1}{l|}{\begin{tabular}[c]{@{}l@{}}$O_{26}:$ $(-)^4$\\ $O_{27},O_{28}:$ $(-)^3+$\\ $O_{29},O_{30}:$ $(-)^2(+)^2$\end{tabular}} \\ \hline
		\multicolumn{1}{|l|}{\multirow{3}{*}{6 evanescent}} & \multirow{2}{*}{5 C-even}  & 3 length-4  &     \multicolumn{1}{l|}{$O_{31},O_{32},O_{33}$}                                                                                                                     \\ \cline{3-4}
		\multicolumn{1}{|l|}{}                              &                            & 2 length-5  &            \multicolumn{1}{l|}{$O_{34},O_{35}$}                                                                                                               \\ \cline{2-4}
		\multicolumn{1}{|l|}{}                              & 1 C-odd                    & 1 length-2  &                       \multicolumn{1}{l|}{$O_{36}$}                                                                                                      \\ \cline{1-4}
	\end{tabular}
\caption{The arrangement of the dimension-10 single-trace operator basis. The subscript $i$ in $O_i$ corresponds to the row and column in the $Z$ matrix. We first divide the operators into the physical sector and the evanescent sector. Within each sector, the operators are further classified according to the C-parities, the lengths and the helicities sector gradually. }
\label{opertable}
\end{table}

\section{Renormalization}\label{sec:renorm}

In this section, we show some details of renormalization. We first discuss the structure of the $Z$ matrix under our arrangement of operators in Section~\ref{sec:Zform}. Then in Section~\ref{sec:strategy} we show the divergence structure of form factors and the strategy for renormalization. In Section~\ref{sec:scheme}, we discuss the $Z$ matrix in two different renormalization schemes: the $\overline{\text{MS}}$ scheme \cite{Bardeen:1978yd} and the finite renormalization scheme \cite{Buras:1989xd,Bondi:1989nq,Dugan:1990df}.

\subsection{Structure of the $Z$ matrix}\label{sec:Zform}

Before showing how to calculate the $Z$ matrix, in this section we first give a description of the blockwise structure of the $Z$ matrix resulting from our operator basis choice in Section~\ref{allopers}. The operators are divided into physical and evanescent operators, and we recall that physical operators refer to those that are non-vanishing in four spacetime dimensions. The $Z$ matrix has the following structure
\begin{align}
\left(
\begin{array}{cc}
Z_{\text{pp}} & Z_{\text{pe}} \\
Z_{\text{ep}} & Z_{\text{ee}}
\end{array}
\right).
\label{Zfinite}
\end{align}
In a subscript, an ``e" denotes ``evanescent" and a ``p" denotes ``physical". For example, the $Z_{\text{pp}}$ denotes the block of physical-to-physical mixing. Other blocks are similarly defined. Since there is no mixing between the C-even and C-odd operators, the $Z$ matrix can be further divided into
\begin{align}
\left(
\begin{array}{cccc}
Z_{\text{pp}}^{\text{even}} & 0 &Z_{\text{pe}}^{\text{even}} & 0\\
0 & Z_{\text{pp}}^{\text{odd}} &0 & Z_{\text{pe}}^{\text{odd}}\\
Z_{\text{ep}}^{\text{even}} & 0 &Z_{\text{ee}}^{\text{even}} & 0\\
0 & Z_{\text{ep}}^{\text{odd}} &0 & Z_{\text{ee}}^{\text{odd}}\\
\end{array}
\right)\,.\label{Zstructure}
\end{align}
Each block can be further arranged according to the lengths of the operators as follows
\begin{align}
&Z_{\text{pp}}^{\text{even}}=\left(
\begin{array}{cccc}
Z_{\text{pp},2\to 2}^{\text{even}} & Z_{\text{pp},2\to 3}^{\text{even}} & Z_{\text{pp},2\to 4}^{\text{even}}& Z_{\text{pp},2\to 5}^{\text{even}}\\
Z_{\text{pp},3\to 2}^{\text{even}} & Z_{\text{pp},3\to 3}^{\text{even}} & Z_{\text{pp},3\to 4}^{\text{even}}& Z_{\text{pp},3\to 5}^{\text{even}}\\
Z_{\text{pp},4\to 2}^{\text{even}} & Z_{\text{pp},4\to 3}^{\text{even}} & Z_{\text{pp},4\to 4}^{\text{even}}& Z_{\text{pp},4\to 5}^{\text{even}}\\
Z_{\text{pp},5\to 2}^{\text{even}} & Z_{\text{pp},5\to 3}^{\text{even}} & Z_{\text{pp},5\to 4}^{\text{even}}& Z_{\text{pp},5\to 5}^{\text{even}}\\
\end{array}
\right)\,,
&&Z_{\text{pp}}^{\text{odd}}=\left(
\begin{array}{cc}
Z_{\text{pp},3\to 3}^{\text{odd}} & Z_{\text{pp},3\to 4}^{\text{odd}}\\
Z_{\text{pp},4\to 3}^{\text{odd}} & Z_{\text{pp},4\to 4}^{\text{odd}}\\
\end{array}
\right)\,,
\label{Zpp}\\
&Z_{\text{pe}}^{\text{even}}=\left(
\begin{array}{cc}
Z_{\text{pe},2\to 4}^{\text{even}}& Z_{\text{pe},2\to 5}^{\text{even}}\\
Z_{\text{pe},3\to 4}^{\text{even}}& Z_{\text{pe},3\to 5}^{\text{even}}\\
Z_{\text{pe},4\to 4}^{\text{even}}& Z_{\text{pe},4\to 5}^{\text{even}}\\
Z_{\text{pe},5\to 4}^{\text{even}}& Z_{\text{pe},5\to 5}^{\text{even}}\\
\end{array}
\right)\,,
&&Z_{\text{pe}}^{\text{odd}}=\left(
\begin{array}{c}
Z_{\text{pe},3\to 4}^{\text{odd}}\\
Z_{\text{pe},4\to 4}^{\text{odd}}\\
\end{array}
\right)\,,
\label{Zpe}\\
&Z_{\text{ep}}^{\text{even}}=\left(
\begin{array}{cccc}
Z_{\text{ep},4\to 2}^{\text{even}} & Z_{\text{ep},4\to 3}^{\text{even}} & Z_{\text{ep},4\to 4}^{\text{even}}& Z_{\text{ep},4\to 5}^{\text{even}}\\
Z_{\text{ep},5\to 2}^{\text{even}} & Z_{\text{ep},5\to 3}^{\text{even}} & Z_{\text{ep},5\to 4}^{\text{even}}& Z_{\text{ep},5\to 5}^{\text{even}}\\
\end{array}
\right)\,,
&&Z_{\text{ep}}^{\text{odd}}=\left(
\begin{array}{cc}
Z_{\text{ep},4\to 3}^{\text{odd}} & Z_{\text{ep},4\to 4}^{\text{odd}}\\
\end{array}
\right)\,,
\label{Zep}\\
&Z_{\text{ee}}^{\text{even}}=\left(
\begin{array}{cc}
Z_{\text{ee},4\to 4}^{\text{even}}& Z_{\text{ee},4\to 5}^{\text{even}}\\
Z_{\text{ee},5\to 4}^{\text{even}}& Z_{\text{ee},5\to 5}^{\text{even}}\\
\end{array}
\right)\,,
&&Z_{\text{ee}}^{\text{odd}}=\left(
\begin{array}{c}
Z_{\text{ee},4\to 4}^{\text{odd}}\\
\end{array}
\right)\,,
\label{Zee}
\end{align}
where $Z_{L\to L'}$ represents the mixing from length-$L$ operators to length-$L'$ operators. The dilatation matrix $\mathcal{D}$, defined in \eqref{overallgamma}, has the same structure as the $Z$ matrix.

The anomalous dimensions are the eigenvalues of $\mathcal{D}$, which are given by the equation
\begin{align}
	\text{Det}\left(\mathcal{D}-\boldsymbol{1}\ \gamma\right)=0\,,
	\label{eigen}
\end{align}
where Det$(M)$ means the determinant of the matrix $M$ and $\boldsymbol{1}$ denotes the identity matrix. $\gamma$ can be expanded in the coupling constant as
\begin{align}
	\gamma=\sum_{l=1}\left(\frac{\alpha_s}{4\pi}\right)^l\gamma^{(l)}\,.
	\label{gammaexpand}
\end{align}
In this work, we give a calculation of the anomalous dimensions up to two loops. At the one-loop order, an operator would not mix with operators of lower lengths. In other words, $Z^{(1)}$ is a block upper triangular matrix according to the lengths of operators. Based on this fact, it is not hard to derive that the calculation of two-loop anomalous dimensions only requires the blocks $Z^{(1)}_{L\to L}$, $Z^{(1)}_{L\to L+1}$, $Z^{(2)}_{L\to L-1}$ and $Z^{(2)}_{L\to L}$.

\subsection{Divergence structure of form factors}\label{sec:strategy}
{Form factors of bare operators (\emph{i.e.}~bare form factors) involve both the ultraviolet (UV) divergences and the the infrared (IR) divergences. The $Z$ matrix is the renormalization matrix to absorb the UV divergences and can be calculated by requiring that renormalized form factors have only IR divergences. Below we give a detailed discussion about the divergences and renormalization of form factors. The computation of bare form factors will be the topic of the next section. }

According to \eqref{renormoperator}, the form factor of an renormalized operator $O_i$ reads
\begin{align}
\mathcal{F}_i=Z_i^{\ j} \mathcal{F}_{j,\text{b}}\,,
\label{Frenorm}
\end{align}
where $\{\mathcal{F}_{j,\text{b}}\}$ are the bare form factors. The $n$-point bare form factors can be expanded as
\begin{align}
\mathcal{F}_{i,\text{b}}=g_0^{\delta_n} \sum_{l=0} \left(\frac{\alpha_0}{4\pi}\right)^{l}\mathcal{F}_{i,\text{b}}^{(l)}\,,
\label{bare}
\end{align}
with $\alpha_0=\frac{g_0^2}{4\pi}$ the bare coupling constant and $\delta_n$ equal $(n-2)$ according to the convention that each monomial operator has a factor $(-\text{i}g_0)^{L-2}$ as mentioned in Section~\ref{ymnotation}. In the $\overline{\text{MS}}$ scheme, the two-loop renormalization of $\alpha_0$ reads (see {\emph{e.g.}~\cite{Gehrmann:2011aa})
\begin{align}
\alpha_0=\alpha_s S_\epsilon^{-1}\mu^{2\epsilon}\left(1-\frac{\beta_0}{\epsilon}\frac{\alpha_s}{4\pi}
+\left(\frac{\beta_0^2}{\epsilon^2}-\frac{\beta_1}{2\epsilon}\right)\left(\frac{\alpha_s}{4\pi}\right)^2+\mathcal{O}(\alpha_s^3)\right)\,,
\label{renormalpha}
\end{align}
where $\alpha_s=\frac{g_s^2}{4\pi}$ is the renormalized coupling constant, $S_\epsilon=(4\pi e^{\gamma_E})^\epsilon$, and $\mu$ is the renormalization scale. The constants in \eqref{renormalpha} are
\begin{align}\label{eq:beta0beta1}
\beta_0=\frac{11N_c}{3},\quad \beta_1=\frac{34N_c^2}{3}\,.
\end{align}
The $Z$ matrix can be expanded as
\begin{align}
Z^{\ j}_i=\delta^j_i+\sum_{l=1}\left(\frac{\alpha_s}{4\pi}\right)^l{{Z^{(l)}}^{\ j}_i}\,.
\label{Zexpand}
\end{align}
Substituting \eqref{Zexpand} into \eqref{overallgamma}, together with the beta function in the pure Yang-Mills theory:
\begin{align}
 \mu\frac{\text{d}\alpha_s}{\text{d}\mu}=-2\epsilon\alpha_s-\frac{\beta_0}{2\pi}\alpha_s^2+\mathcal{O}(\alpha_s^3)\,,
\end{align}
one obtains the expansion of the dilatation matrix as
\begin{align}
&\mathcal{D}^{(1)}=2\epsilon Z^{(1)}\,,\label{gamma1}\\
&\mathcal{D}^{(2)}=4\epsilon Z^{(2)}-2\epsilon \left(Z^{(1)}\right)^2+2\beta_0Z^{(1)}\,.
\label{gamma2}
\end{align}
According to \eqref{gamma2}, a finite two-loop dilatation matrix requires the following matrix equation
\begin{align}
	Z^{(2)}|_{\frac{1}{\epsilon^2}-\text{part}}=\Big(\frac{1}{2}(Z^{(1)})^2-\frac{\beta_0}{2\epsilon}Z^{(1)}\Big)|_{\frac{1}{\epsilon^2}-\text{part}}\,.
\label{z2ep2ms}
\end{align}

{
A renormalized form factor can be expanded in the renormalized coupling $\alpha_s$ in a form similar to \eqref{bare},
}
\begin{align}
\mathcal{F}_i=g_s^{\delta_n}S_\epsilon^{-\frac{\delta_n}{2}}\sum_{l=0} \left(\frac{\alpha_s}{4\pi}\right)^{l}\mathcal{F}_i^{(l)}\,,
\end{align}
{
The relations between ${\cal F}_i$ and ${\cal F}_{i,{\rm b}}$ can be obtained using (3.9), (3.11) and (3.13), and up to the two-loop order, they are given by
}
\begin{align}
&\mathcal{F}_i^{(0)}=\mathcal{F}_{i,\text{b}}^{(0)}\,,\label{reF0}\\
&\mathcal{F}_i^{(1)}=S_\epsilon^{-1}\mathcal{F}_{i,\text{b}}^{(1)}+\left({Z^{(1)}}_i^{\ j}-\frac{\delta_n}{2}\frac{\beta_0}{\epsilon}\delta^{\ j}_i\right)\mathcal{F}_{j,\text{b}}^{(0)}\,,\label{reF1}\\
&\mathcal{F}_i^{(2)}=S_\epsilon^{-2}\mathcal{F}_{i,\text{b}}^{(2)}+S_\epsilon^{-1}\left({Z^{(1)}}^{\ j}_i-\left(1+\frac{\delta_n}{2}\right)\frac{\beta_0}{\epsilon}\delta^{\ j}_i\right)\mathcal{F}_{j,\text{b}}^{(1)}\nonumber\\
&\quad \qquad+
\left({Z^{(2)}}^{\ j}_i-\frac{\delta_n}{2}\frac{\beta_0}{\epsilon}{Z^{(1)}}^{\ j}_i+\frac{\delta_n^2+2\delta_n}{8}\frac{\beta_0^2}{\epsilon^2}\delta^{\ j}_i-\frac{\delta_n}{4}\frac{\beta_1}{\epsilon}\delta^{\ j}_i\right)\mathcal{F}_{j,\text{b}}^{(0)}\,,
\label{reF2}
\end{align}

{
A renormalized form factor only involves the IR divergences. The IR divergences is well understood, and up to two loops, the IR divergences of renormalized form factors take the universal form as \cite{Catani:1998bh}}
\begin{align}
&\mathcal{F}_i^{(1)}\big{|}_{\text{div}}=\mathcal{F}_i^{(1)}\big{|}_{\text{IR}}=I^{(1)}(\epsilon)\mathcal{F}_i^{(0)}\label{ir1} \,,\\
&\mathcal{F}_i^{(2)}\big{|}_{\text{div}}=\mathcal{F}_i^{(2)}\big{|}_{\text{IR}}=I^{(2)}(\epsilon)\mathcal{F}_i^{(0)}+I^{(1)}(\epsilon)\mathcal{F}_i^{(1)}\,,
\label{ir2}
\end{align}
The factors $I^{(1)}$ and $I^{(2)}$ read}
\begin{align}
I_{n}^{(1)}(\epsilon) &= - {e^{\gamma_E \epsilon} \over \Gamma(1-\epsilon)} \bigg( \frac{N_c}{\epsilon^2} + \frac{\beta_0}{2 \epsilon} \bigg) \sum_{i=1}^n (-{s_{i,i+1}} )^{-\epsilon} \,, \label{thei1} \\
I_{n}^{(2)}(\epsilon) &= - {1\over2} \big[ I^{(1)}(\epsilon) \big]^2   -  {\beta_0 \over \epsilon} I^{(1)}(\epsilon)
+ {e^{-\gamma_E \epsilon} \Gamma(1-2\epsilon) \over \Gamma(1-\epsilon)} \left[ \frac{\beta_0}{\epsilon} + {\cal K}\right] I^{(1)}(2\epsilon)  + n {e^{\gamma_E \epsilon} \over \epsilon \Gamma(1-\epsilon)}{\cal H}_{\Omega,g}^{(2)}  \,, \nonumber
\end{align}
with
\begin{align}
{\cal K} = \left({67\over9} - {\pi^2\over3}\right) N_c\,, \qquad
{\cal H}_{\Omega,g}^{(2)}  =  \left( \frac{\zeta_3}{2} + {5\over12} + {11\pi^2 \over 144} \right)N_c^2   \,.
\end{align}
{By requiring that \eqref{reF1} and \eqref{reF2} have the same divergences as \eqref{ir1} and \eqref{ir2}, namely
\begin{equation}
\mathcal{F}_i^{(\ell)} - \mathcal{F}_i^{(1)}\big{|}_{\text{IR}} = {\cal O}(\epsilon^0)\,, \nonumber
\end{equation}
one can build linear equations between ${Z^{(l)}}$'s and bare form factors. The ${\cal O}(\epsilon^0)$ term depends on the renormalization scheme. For example, the linear equations for the $\overline{\text{MS}}$ scheme are given in \eqref{msbareqa1} and \eqref{msbareqa2}.}

\subsection{$Z$ matrix and renormalization schemes}\label{sec:scheme}
The definition of the $Z$ matrix depends on the choice of renormalization schemes. 
In this subsection, we discuss $Z$ matrix in the $\overline{\text{MS}}$ scheme \cite{Bardeen:1978yd} and the finite renormalization scheme \cite{Buras:1989xd,Bondi:1989nq,Dugan:1990df} in Section~\ref{zms} and Section~\ref{zfin} respectively. 
We will show that it is easier to compute the physical anomalous dimensions in the finite renormalization scheme.

\subsubsection{$\overline{\text{MS}}$ scheme}\label{zms}

In the $\overline{\text{MS}}$ scheme, the $Z$ matrix is determined only by the UV divergences of form factors.
According to \eqref{reF0}-\eqref{ir2}, the relations between $Z$ matrix and bare form factors reads
\begin{align}
{Z^{(1)}}^{\ j}_i\mathcal{F}_{j,\text{b}}^{(0)}&=\left(I^{(1)}(\epsilon)\mathcal{F}_{i,\text{b}}^{(0)}-S_\epsilon^{-1}\mathcal{F}_{i,\text{b}}^{(1)}+\frac{\delta_n}{2}\frac{\beta_0}{\epsilon}\mathcal{F}_{i,\text{b}}^{(0)}\right)\bigg{|}_{\text{divergent part}}\,,\label{msbareqa1}\\
{Z^{(2)}}^{\ j}_i\mathcal{F}_{j,\text{b}}^{(0)}&=\left[I^{(2)}(\epsilon)\mathcal{F}_{i,\text{b}}^{(0)}+I^{(1)}(\epsilon)\mathcal{F}_i^{(1)}-S_\epsilon^{-2}\mathcal{F}_{i,\text{b}}^{(2)}-S_\epsilon^{-1}\left({Z^{(1)}}^{\ j}_i-\left(1+\frac{\delta_n}{2}\right)\frac{\beta_0}{\epsilon}\delta^{\ j}_i\right)\mathcal{F}_{j,\text{b}}^{(1)}\right.\nonumber\\
& \qquad -
\left.\left(-\frac{\delta_n}{2}\frac{\beta_0}{\epsilon}{Z^{(1)}}^{\ j}_i+\frac{\delta_n^2+2\delta_n}{8}\frac{\beta_0^2}{\epsilon^2}\delta^{\ j}_i-\frac{\delta_n}{4}\frac{\beta_1}{\epsilon}\delta^{\ j}_i\right)\mathcal{F}_{j,\text{b}}^{(0)}\right]\bigg{|}_{\text{divergent part}}\,.
\label{msbareqa2}
\end{align}
We make two remarks on practical computations.
In our problem, the rank of the $Z$ matrix is finite, thus one can reconstruct the $Z$ matrix from a sufficient set of numerical points of equations \eqref{msbareqa1} and \eqref{msbareqa2}. In practice, we would assign each Lorentz invariant a random numerical value during the calculation of bare form factors as will be seen in Section~\ref{calcbare} (there is no linear relation between Lorentz invariants in general $d$ spacetime dimensions). 
As another remark, since an operator may mix to operators of different lengths, it is necessary to consider the renormalization of form factors with different numbers of external gluons. A convenient order to renormalize the form factors of a given operator is from lower-point ones to higher-point ones. In this way, one can use the mixing matrix elements associated with lower-length operators as input in the renormalization of a higher-point form factor. This not only simplifies the computation but also provides a check for the computation.

In the $\overline{MS}$ scheme, the $Z$ matrices have the blockwise structures
\begin{align}
\left(
\begin{array}{cc}
Z_{\text{pp}}^{(1)} & Z_{\text{pe}}^{(1)} \\
0 & Z_{\text{ee}}^{(1)}
\end{array}
\right), \qquad 
\left(
\begin{array}{cc}
Z_{\text{pp}}^{(2)} & Z_{\text{pe}}^{(2)} \\
Z_{\text{ep}}^{(2)} & Z_{\text{ee}}^{(2)}
\end{array}
\right),
\label{Zfinite12}
\end{align}
where the block $Z_{\text{ep}}^{(1)}$ vanishes. This is due to the fact that the fact that the form factor of an evanescent operator is one order higher in the $\epsilon$ expansion. According to \eqref{gamma1} and \eqref{gamma2}, the dilatation matrices have similar blockwise structures
\begin{align}
\left(
\begin{array}{cc}
\mathcal{D}_{\text{pp}}^{(1)} & \mathcal{D}_{\text{pe}}^{(1)} \\
0 & \mathcal{D}_{\text{ee}}^{(1)}
\end{array}
\right) , \qquad
\left(
\begin{array}{cc}
\mathcal{D}_{\text{pp}}^{(2)} & \mathcal{D}_{\text{pe}}^{(2)} \\
\mathcal{D}_{\text{ep}}^{(2)} & \mathcal{D}_{\text{ee}}^{(2)}
\end{array}
\right) .
\label{dilatationms}
\end{align}
Note that starting from two loops, all four blocks of $Z$ and $\mathcal{D}$ matrices are in general non-vanishing.

\subsubsection{Finite renormalization scheme}\label{zfin}
We give an introduction for the finite renormalization scheme in this subsection. To distinguish from the $\overline{\text{MS}}$ scheme, we use $\hat{Z}$ and $\hat{\mathcal{D}}$ to denote the $Z$ matrix and dilatation matrix in the finite renormalization scheme. 
The most important feature of the finite renormalization scheme is that {the block $\hat{\mathcal{D}}_{\text{ep}}^{(l)}$ in the dilatation matrix is $\mathcal{O}(\epsilon)$ at all orders. Therefore in the $\epsilon\to 0$ limit, the dilatation matrix take the block upper triangular form \cite{Buras:1989xd,Bondi:1989nq,Dugan:1990df}:}
\begin{align}
\left(
\begin{array}{cc}
\hat{\mathcal{D}}_{\text{pp}}^{(l)} & \hat{\mathcal{D}}_{\text{pe}}^{(l)} \\
0 & \hat{\mathcal{D}}_{\text{ee}}^{(l)}
\end{array}
\right).
\label{uptriangulargamma}
\end{align}
We will see that this simplifies the calculation of physical anomalous dimensions.

In the finite renormalization scheme, the renormalization of physical operators are the same as the ones in the $\overline{\text{MS}}$ scheme and we have
\begin{align}\label{zppsame}
	\hat{Z}^{(l)}_{\text{pp}}=Z^{(l)}_{\text{pp}},\qquad
 \hat{Z}^{(l)}_{\text{pe}}=Z^{(l)}_{\text{pe}}\,.
\end{align}
While the renormalization of evanescent operators is different.
This scheme takes into account the fact that the form factor of an evanescent operator is one order higher in the $\epsilon$ expansion, and
the mixing from evanescent to physical operators is finite and should also be subtracted. 
In other words, one will modify the RHS of \eqref{msbareqa1}-\eqref{msbareqa2} by taking into account some ``finite'' part of form factors.
Below we first describe how we compute the $Z$ matrix in this scheme, and then we explain how it produces the wanted form of dilatation matrix as \eqref{uptriangulargamma}.

Given an evanescent operator $O_i$, we will separate ${\hat{Z}}^{(l)\ j}_{\quad  i}$ as two parts:
\begin{align}\label{finZ2part}
{\hat{Z}}^{(l)\ j}_{\quad  i} = {\hat{Z}}^{(l)\ j}_{\quad  i}\big|_{\text{div}} + {\hat{Z}}^{(l)\ j}_{\quad  i}\big|_{\text{fin}}\,.
\end{align}
The calculation of the divergent part ${\hat{Z}}^{(l)\ j}_{\quad  i}\big|_{\text{div}}$ is similar to that of the $Z$ matrix in the $\overline{\text{MS}}$ scheme, and the calculation of the finite part ${\hat{Z}}^{(l)\ j}_{\quad  i}\big|_{\text{fin}}$ will be discussed in detail below.
In the following, the index $i$ in ${\hat{Z}}^{(l)\ j}_{\quad  i}$ will refer only to the evanescent operators while $j$ can refer to both physical and evanescent operators. 

Consider first the one-loop order. We have
\begin{align}\label{1loopsame}
{\hat{Z}}^{(1)\ j}_{\quad  i}\big|_{\text{div}}={Z}^{(1)\ j}_{\quad i}\,,
\end{align}
where ${Z}^{(1)}$ is the ${\overline{\text{MS}}}$ $Z$ matrix.
To compute the finite part ${\hat{Z}}^{(1)\ j}_{\quad i}\big|_{\text{fin}}$, one can consider the one-loop form factors at 4-dimensional numerical points (for example, using the spinor helicity formalism). Since the tree-level form factors of evanescent operators vanish in 4-dimensional spacetime, only the parts proportional to physical operators remains in the form factor. To subtract also the finite evanescent-to-physical mixing, the one-loop formula \eqref{msbareqa1} is modified as
\begin{align}
{{\hat{Z}}'^{(1)\ j}_{\quad  i}}\mathcal{F}_{j,\text{b}}^{(0),\text{4$d$}}&=-S_\epsilon^{-1}\mathcal{F}_{i,\text{b}}^{(1),\text{4$d$}}\big{|}_{\text{divergent and finite parts}}\,,\label{fineqa1}
\end{align}
and we have
\begin{align}\label{zhat}
{\hat{Z}}^{(1)\ j}_{\quad  i}\big|_{\text{fin}} = \text{finite part of }{\hat{Z}}'^{(1)\ j}_{\quad i}\,.
\end{align}
The superscript ``4$d$" means the form factor is calculated in 4-dimensional numerical points. It should be clear that $j$ in \eqref{zhat} can only refer to physical operators, thus ${\hat{Z}}^{(1)\ j}_{\quad  i}\big|_{\text{fin}}$ only contribute to the block ${\hat Z}_{\text{ep}}^{(1)}$.

Next we consider the two-loop order. 
The calculation of ${\hat{Z}}^{(2)\ j}_{\quad i}\big|_{\text{div}}$ is similar to the $\overline{\text{MS}}$ scheme, except that we need to replace all $Z^{(1)}$ in \eqref{msbareqa2} by $\hat{Z}^{(1)}$. Since $\hat{Z}^{(1)}\neq Z^{(1)}$,  at the two-loop order one does not have the simple relation similar to \eqref{1loopsame}. For the finite part, we consider similarly the form factors at 4-dimensional numerical points and calculate ${\hat{Z}'^{(2)\ j}}_{\quad i}$ via the formula:
\begin{align}
{\hat{Z}'^{(2)\ j}}_{\quad i}\mathcal{F}_{j,\text{b}}^{(0),\text{4$d$}}&=\left[I^{(1)}(\epsilon)\mathcal{F}_i^{(1),\text{4$d$}}-S_\epsilon^{-2}\mathcal{F}_{i,\text{b}}^{(2),\text{4$d$}}-S_\epsilon^{-1}\left({\hat{Z}^{(1)\ j}}_{\quad i}-\left(1+\frac{\delta_n}{2}\right)\frac{\beta_0}{\epsilon}\delta^{\ j}_i\right)\mathcal{F}_{j,\text{b}}^{(1),\text{4$d$}}\right.\nonumber\\
&\qquad +
\left.\frac{\delta_n}{2}\frac{\beta_0}{\epsilon}{\hat{Z}^{(1)\ j}}_{\quad i}
\mathcal{F}_{j,\text{b}}^{(0),\text{4$d$}}\right]\bigg{|}_{\text{divergent and finite parts}}\,,
\label{fineqa2}
\end{align}
and the two-loop finite part is given by
\begin{align}\label{zhat2}
{\hat{Z}}^{(2)\ j}_{\quad  i}\big|_{\text{fin}} = \text{finite part of }{\hat{Z}}'^{(2)\ j}_{\quad i}\,.
\end{align}
Again, ${\hat{Z}}^{(2)\ j}_{\quad  i}\big|_{\text{fin}}$ only contributes to the block $Z_{\text{ep}}^{(2)}$.
At the two-loop order, one can check the divergent mixing to the physical operators in ${\hat{Z}}'^{(2)\ j}_{\quad i}$ should be the same as in ${\hat{Z}}^{(2)\ j}_{\quad i}\big|_{\text{div}}$. From \eqref{gamma2}, one can see that ${\hat{Z}}^{(2)\ j}_{\quad  i}\big|_{\text{fin}}$ contributes to the $\mathcal{O}(\epsilon)$ of the two-loop dilatation matrix, therefore it does not contribute to the calculation of the two-loop anomalous dimensions (it begins to contribute at the three-loop order). 

As mentioned at the beginning of this subsection, the key feature of the finite renormalization scheme is that the dilatation matrix has the form of \eqref{uptriangulargamma} {under the limit $\epsilon\to 0$}.
{At the one-loop order, this is straightforward since ${\hat{Z}^{(1)}}_{\text{ep}}$ is finite, thus $\hat{\mathcal{D}}^{(1)}_{\text{ep}}\sim{\cal O}(\epsilon)$. 
At the two-loop order, the leading divergence of ${\hat{Z}^{(2)}}_{\text{ep}}$ is of ${\cal O}(1/\epsilon)$, and the relation \eqref{z2ep2ms} still applies to the leading divergences (which is one order higher in the $\epsilon$-expansion than usual cases) \cite{tHooft:1972tcz,Dugan:1990df}: 
\begin{align}
{\hat{Z}}_{\text{ep}}^{(2)}|_{\frac{1}{\epsilon}-\text{part}}=\frac{1}{2} \big( {\hat{Z}}_{\text{ep}}^{(1)}{\hat{Z}}_{\text{pp}}^{(1)}+{\hat{Z}}_{\text{ee}}^{(1)}{\hat{Z}}_{\text{ep}}^{(1)} \big) -\frac{\beta_0}{2\epsilon}{\hat{Z}}_{\text{ep}}^{(1)} \,.
\label{vansih2loop}
\end{align}
Using \eqref{vansih2loop} and \eqref{gamma2}, it should then be clear that $\hat{\mathcal{D}}^{(2)}_{\text{ep}}$ is also $\mathcal{O}(\epsilon)$.} Note that the block ${\hat{Z}}_{\text{ep}}^{(1)}$ which is finite is necessary in this cancellation.
We check that our explicit two-loop calculations indeed confirm this structure. 

Since block upper triangular form \eqref{uptriangulargamma}, the physical anomalous dimensions are just the eigenvalues of the $\hat{\mathcal{D}}_{\text{pp}}$. This does not mean that evanescent operators have no effect on the physical anomalous dimensions. At the two-loop order, the effect of the evanescent operators on $\hat{\mathcal{D}}^{(2)}_{\text{pp}}$ comes from the term $(-2\epsilon {\hat{Z}}^{(1)}_{\text{pe}}{\hat{Z}}^{(1)}_{\text{ep}})$ according to \eqref{gamma2}. Obviously, evanescent operators should be renormalized up to the one-loop order in the calculation of the two-loop physical anomalous dimensions. 

We point out here that anomalous dimensions are scheme dependent, due to the non-vanishing beta function in the pure YM theory, and therefore, the results in the finite renormalization scheme are different from the ones in the $\overline{\text{MS}}$ scheme. On the other hand, at the conformal fixed point, anomalous dimensions should be independent of the renormalization scheme. A detailed discussion of the scheme dependence of anomalous dimensions will be given in Section~\ref{sec:fixed point}.

\section{Calculation of bare form factors}\label{calcbare}

In this section, we consider the computation of bare form factors up to the two-loop order. In Section~\ref{allcuts}, we give an overall description of our calculation. 
In Section~\ref{tensorreduce}, we discuss two methods for integral tensor reduction in detail.

\subsection{Unitarity-IBP method}\label{allcuts}

The main strategy of our calculation is based on a combination of the unitarity method \cite{Bern:1994cg,Bern:1994zx,Britto:2004nc} and the IBP reduction \cite{Chetyrkin:1981qh,Tkachov:1981wb}. 
This strategy has been applied to compute form factors (and Higgs amplitudes) in \cite{Jin:2018fak, Jin:2019ile, Jin:2019opr} and for pure gluon amplitudes in \cite{Boels:2017gyc, Boels:2018nrr, Jin:2019nya}.
The numerical IBP method by cuts was also studied in \cite{Kosower:2011ty,Larsen:2015ped,Ita:2015tya,Georgoudis:2016wff,Abreu:2017hqn,Abreu:2017xsl}.

A loop form factor can be written as a linear combination of a set of IBP master integrals as
\begin{align}
\mathcal{F}^{(l)}=\sum_i c_i I^{(l)}_i\,,
\label{Fdecomposition}
\end{align}
where the coefficients $c_i$ are rational functions of the external Lorentz invariants and the spacetime dimension $d$. If one imposes a unitarity cut on \eqref{Fdecomposition}, one gets
\begin{align}
\mathcal{F}^{(l)}|_{\text{cut}}=\sum_{i'} c_{i'} I^{(l)}_{i'}|_{\text{cut}}\,,
\label{cutdecomposition}
\end{align}
where the sum of $i'$ runs over all master integrals which can be detected by the cut, \emph{i.e.} non-vanishing under the cut condition. Below we show how to calculate the coefficients $c_{i'}$ in a single cut. In the end, we need to choose a set of cuts to cover all the masters and then calculate their coefficients.

Given a cut, we first compute the cut-integrand of a loop form factor as the product of tree form factor and amplitudes:
\begin{align}
\mathcal{F}^{(l)}|_{\text{cut}}=\sum_{\text{helicities}} \mathcal{F}^{(0)} \times(\prod_i \mathcal{A}^{(0)}_i)\,,
\label{cutF}
\end{align}
where $\mathcal{F}^{(0)}$ denotes a tree form factor and $\mathcal{A}^{(0)}_i$ denote tree scattering amplitudes. Since we use $d$-dimensional unitarity cuts, the tree-level results are obtained in terms of Lorentz product of $d$-dimensional momenta $\{p_i, l_a\}$ and polarization vectors $e_i$, and compact formula for form factors can be found in Appendix F of \cite{Jin:2022ivc}.
Our results will be obtained in the CDR scheme which is valid for states in general $d$ spacetime dimensions.\footnote{
We mention that there are some alternative dimensional reduction schemes where the $4$-dimensional gauge fields may be expressed as $D$-dimensional components plus the $\epsilon$-scalars \cite{Siegel:1979wq, Capper:1979ns, Jack:1993ws, Harlander:2006rj, Nandan:2014oga}.
One may also consider 6-dimensional spinor helicity formalism for form factors as in \cite{Huber:2019fea}. The operator renormalization has also been considered for gauge theories in six and eight spacetime dimensions \cite{Gracey:2015xmw, Gracey:2017nly}.
For the two-loop renormalization involving fermionic evanescent operators and $\gamma_5$, see also \cite{Buras:1989xd, Schubert:1988ke}.
}

The sum of helicities is performed for the polarization vectors $e_l$ of internal cut gluon legs, for which we adopt the $d$-dimension helicity sum:
\begin{align}
e^\mu_{l}  \circ {e^\nu_{l}}^*\equiv \sum_{\text{helicities}} e^\mu_{l}  {e^\nu_{l}}^*=\eta^{\mu\nu}-\frac{q^\mu l^\nu+l^\mu q^\nu}{l\cdot q}\,,
\label{helsum}
\end{align}
where $q$ is a light-like reference momentum. 

After this step,  the cut integrand is given as a rational function of Lorentz invariants. In particular, it includes the Lorentz products of the loop momenta and external polarization vectors like $l_j \cdot e_i$ which can not be expanded in terms of propagators, and thus the IBP reduction can not be used directly. To eliminate such Lorentz invariants, we multiply back the cut propagators in the cut integrand and then do tensor reduction for Feynman integrals. 
We adopt two different methods for tensor reduction: 1) the gauge invariant basis projection method, and 2) a hybrid method combing loop momentum decomposition and the PV reduction. The first method is efficient in the calculations of 2-point and 3-point form factors, while for form factors with more external gluons, the second method is preferable. We will give a detailed description of these two methods in Section~\ref{tensorreduce}. 

After tensor reduction, we can expand the integrals with a set of chosen propagators which are ready to do the IBP reduction (with the cut condition imposed), and in this work we use the package FIRE6 \cite{Smirnov:2019qkx}. 
After the IBP reduction, the cut form factor is transformed into the desired form as shown in \eqref{cutdecomposition} and one gets $c_{i'}$. 
Due to the complexity of the expressions, the numerical assignment for external Lorentz invariants is also used during the IBP reduction for the case of length-4 and length-5 operators.

To illustrate the above strategy in a concrete setup, below we show all kinds of form factors calculated in this work, as well as all the IBP master integrals and the cuts to cover them.

\begin{figure}[t]
	\centering
	\includegraphics[scale=0.5]{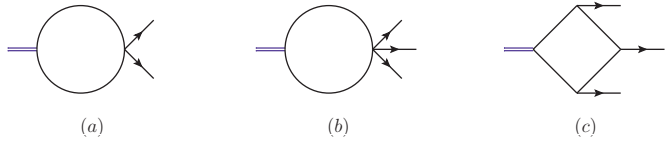}
	\caption{The one-loop masters.}
	\label{1loop_mi}
\end{figure}

\begin{figure}[t]
	\centering
	\includegraphics[scale=0.5]{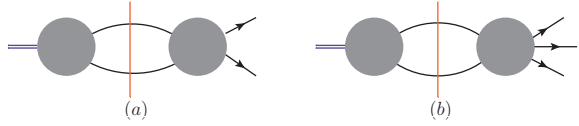}
	\caption{The one-loop cuts.}
	\label{1loop_cuts}
\end{figure}

\begin{figure}[t]
	\centering
	\includegraphics[scale=0.5]{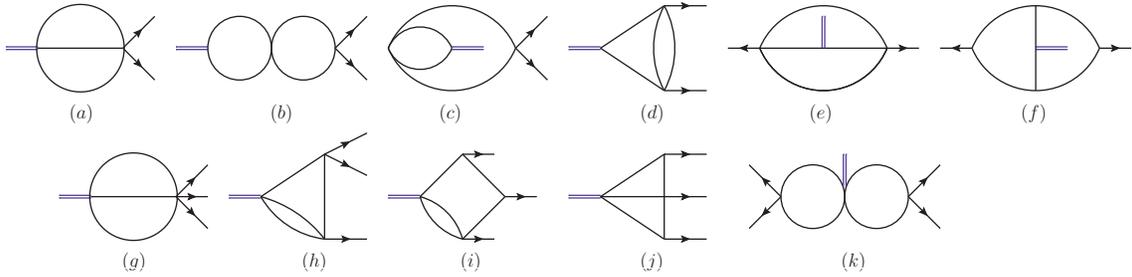}
	\caption{The two-loop masters. The flipped versions of $(h)$ and $(i)$ are also included, which we do not draw.}
	\label{2loop_mi}
\end{figure}

\begin{figure}[t]
	\centering
	\includegraphics[scale=0.5]{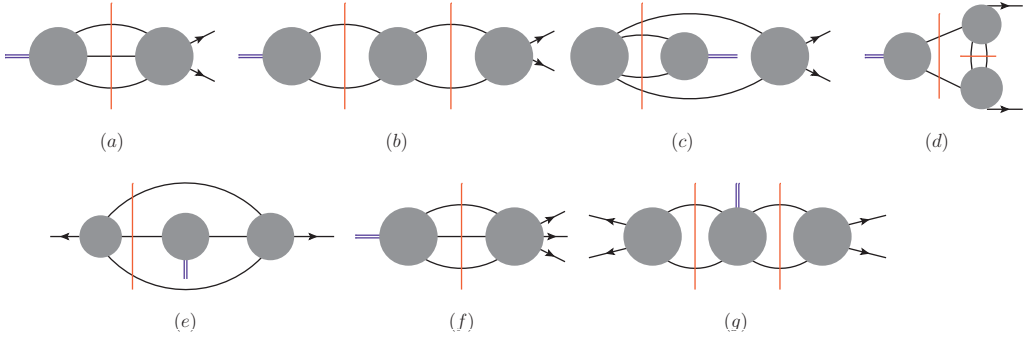}
	\caption{The two-loop cuts}
	\label{2loop_cuts}
\end{figure}

As discussed in Section~\ref{sec:strategy}, we only need to consider the renormalization matrices $Z^{(1)}_{L\to L}$, $Z^{(1)}_{L\to L+1}$, $Z^{(2)}_{L\to L-1}$ and $Z^{(2)}_{L\to L}$. Consequently, the necessary form factors are
\begin{align}
	&\mathcal{F}^{(1)}_{L\to L},\ L=2,3,4,5\,,\qquad \mathcal{F}^{(1)}_{L\to L+1},\ L=2,3,4\,,\\
	&\mathcal{F}^{(2)}_{L\to L},\ L=2,3,4,5\,,\qquad \mathcal{F}^{(2)}_{L\to L-1},\ L=3,4,5\,,
\end{align}
where $\mathcal{F}_{L\to n}$ represents an $n$-point form factor of a length-$L$ operator. For all these form factors, the one-loop and two-loop master integrals are shown in Figure~\ref{1loop_mi} and Figure~\ref{2loop_mi}. And we present all corresponding cuts in Figure~\ref{1loop_cuts} and Figure~\ref{2loop_cuts}. Notice that all external legs outside the loop are included in the double line. Since the calculation is in the large $N_c$ limit, all the cuts are planar and all the tree blocks are color-ordered. The relations between the form factors, master integrals and the cuts are as follows.
\begin{itemize}
	\item The only master for one-loop minimal form factors is $(a)$ in Figure~\ref{1loop_mi}, detected by the cut $(a)$ in Figure~\ref{1loop_cuts}.
	\item The masters for one-loop next-to-minimal form factors are $(a)\sim(c)$ in Figure~\ref{1loop_mi}, detected by the cuts $(a)$ and $(b)$ in Figure~\ref{1loop_cuts} respectively.
	\item A two-loop sub-minimal form factor only has the master $(a)$ in Figure~\ref{2loop_mi}, detected by the cut $(a)$ in Figure~\ref{2loop_cuts}.
	\item A two-loop $2\to2$ form factor includes the masters $(a)\sim(f)$ in Figure~\ref{2loop_mi}. The detecting cuts are $(a)\sim(e)$ in Figure~\ref{2loop_cuts}. Note that since the local operator is a color singlet, the non-planar masters $(c)$, $(e)$ and $(f)$ in Figure~\ref{2loop_mi} are of leading $N_c$ order.
	\item A two-loop $3\to3$ form factor includes the masters $(a),\ (b),\ (d),\ (g) \sim (j)$ in Figure~\ref{2loop_mi}. The flipped version of $(h)$ and $(i)$, which we do not draw, are also included in the masters. The cuts are $(a),\ (b),\ (d)$ and $(f)$ in Figure~\ref{2loop_cuts}.
	\item All two-loop minimal form factors with more than 3 external legs have all the masters of a two-loop $3\to 3$ form factor and the master $(k)$ in Figure~\ref{2loop_mi}. Accordingly, one needs one more cut, which is the cut $(g)$ in Figure~\ref{2loop_cuts}.
\end{itemize}
It is worthwhile noting that coefficients of some master integrals can be calculated via different cuts, \emph{e.g.} the master integral $(j)$ in Figure~\ref{2loop_mi} can be detected by the cuts $(a)$ and $(f)$ in Figure~\ref{2loop_cuts}. In that case, coefficients of the master integral calculated from different cuts must be the same. All two-loop master integrals can be found in \cite{Gehrmann:2000zt,Gehrmann:2001ck}.


\subsection{Two methods for tensor reduction}\label{tensorreduce}

In this subsection, we give a detailed description of two methods for tensor reduction. We first introduce the gauge-invariant basis projection (see \emph{e.g.} \cite{Gehrmann:2011aa,Boels:2018nrr}) in Section~\ref{gibp}. This method takes advantage of the gauge invariance of the cut form factors and is powerful for the form factors with a small number of external legs. Then in Section~\ref{PV}, we introduce a hybrid method, including loop momentum decomposition and the PV reduction\cite{Passarino:1978jh,Kreimer:1991wj}. The second method is efficient for the calculation of the two-loop high-point form factors.

\subsubsection{Gauge invariant basis projection}\label{gibp}

In this section, we introduce the gauge invariant basis projection method. 
Taking into account the gauge invariance of the cut form factor, one has the following ansatz
\begin{align}
\mathcal{F}|_{\text{cut}}=\sum_{i}f_i(p,l) B_i(p,e)\,,
\label{giansatz}
\end{align}
with $\{B_i\}$ a complete set of gauge invariant basis involving only the external momenta and the corresponding polarization vectors. The coefficients $f_i$ are functions of loop momenta and external momenta.

For an $n$-point form factor, the basis can be constructed via the gauge invariant building blocks \cite{Boels:2018nrr}
\begin{align}
A_{i;j}&=e_i\cdot p_{j+1}\  p_i\cdot p_{j+2}-e_i\cdot p_{j+2}\ p_i\cdot p_{j+1},&i\leq j\leq i+n-3\,,\\
C_{i,j}&=e_i\cdot e_j\ p_i\cdot p_j-e_i\cdot p_j\ e_j\cdot p_i\,,&i,j=1,\cdots,n\text{ and }i\neq j\,,
\label{AC}
\end{align}
where the cyclic convention $i+n=i$ is adopted. For example, the gauge-invariant basis of 3-point form factors reads
\begin{align}
A_{1;1}A_{2;2}A_{3;3},\ A_{1;1}C_{2,3},\ A_{2;2}C_{1,3},\ A_{3;3}C_{1,2}\,.
\end{align}

Define the inner product
\begin{align}
B_1\circ B_2 \equiv\sum_{\text{helicities}} B_1 B_2\,,
\end{align}
where the expression for the sum of helicities is given in \eqref{helsum}. One can construct the dual basis $\{B^i\}$ as
\begin{align}
B^i=(G_B^{-1})^{ij}B_j\,,
\label{dualB}
\end{align}
where $G_B$ is the inner product matrix of $\{B_i\}$:
\begin{align}
(G_B)_{ij}=B_i\circ B_j\,.
\end{align}
It is straightforward to verify that $B^i\circ B_j=\delta^{\ i}_j$. The $f_i$ in \eqref{giansatz} can be calculated by
\begin{align}
f_i=B^i\circ \mathcal{F}|_{cut}\,.
\end{align}
In general, the matrix $G_B$ may be complicated and it may be hard to calculate the inverse. One can refer to \cite{Boels:2018nrr} for a modified strategy, which does the projection blockwisely. 

To count the number of bases, we first note that the gauge invariant basis can be classified into $[\frac{n}{2}]+1$ classes as
\begin{align}
A^n,A^{n-2}C, A^{n-4}C^2\,,\cdots, A^{n-2k}C^{k},\cdots\,.
\label{theACs}
\end{align}
Together with the counting of $A$ and $C$, one can derive the number of the basis for an $n$-point form factor:
\begin{align}
\sum_{i=0}^{[\frac{n}{2}]}(n-2)^{n-2i}\binom{n}{2i}\frac{\prod_{j=0}^{i-1}\binom{2(i-j)}{2}}{i!}\,.
\label{numofbases}
\end{align}
where $\binom{n}{m}$ is the binomial coefficient and $[x]$ means the maximal integer not larger than $x$. The counting is sensitive to the number of external legs. The numbers of bases are $4$, $43$ and $558$ respectively for 3-point, 4-point and 5-point form factors. The large number of bases makes the projection method not so practicable for high-point form factors.  In our calculation, the projection method is used for 2-point and 3-point form factors.

\subsubsection{Loop momentum decomposition and PV reduction}\label{PV}
In this subsection, we describe a hybrid method for tensor reduction, which combines the loop momentum decomposition and the PV reduction.

Our problem is to do tensor reduction for  integrals as follows
\begin{align}\label{PVproblem}
  \int [\text{d}l]\frac{\prod_k l_{i_k}^{\mu_k}}{\prod_{j}\text{D}_j}\,,
\end{align}
where $[\text{d}l]=\prod_{i}\frac{\text{d}^dl_i}{(2\pi)^d}$ and $\text{D}_j$ denote propagators.

At first, each loop momentum $l_i^{\mu}$ can be decomposed as
\begin{align}
l_{i}^\mu=\sum_{k}c_{i,k}\ p_k^\mu +l_{i,\perp}^\mu\,,
\label{l decompse}
\end{align}
where $p_i$'s are external momenta in the propagators and $c_{i,k}$ is rational function of $l_i\cdot p_k$ and $p_j\cdot p_k$.
The $l_{i,\perp}^\mu$ is defined to be perpendicular to all external momenta. After the decomposition \eqref{l decompse}, we would have a sum of tensor integrals, which have the following form
\begin{align}\label{PVstep2}
  \int [\text{d}l]\frac{X}{\prod_{j}\text{D}_j}\prod_{k'} l_{i_{k'},\perp}^{\mu_{k'}}\,,
\end{align}
where $X$ is a rational function of $l_i\cdot p,p\cdot p\text{ and }p^\mu$. Such tensor integrals can be further reduced via the PV reduction\eqref{PVstep2}. Since the integrals are transverse to all the external momenta	, the only building block is the transverse metric $\eta_\perp^{\mu\nu}$ (see \emph{e.g.} \cite{Kreimer:1991wj,Henn:2014yza}) which is symmetric and has the following properties:
\begin{align}
{p_i}_\mu \eta_\perp^{\mu\nu}=0,\ {l_{i_m,_\perp}}_\mu \eta_\perp^{\mu\nu}=l_{i_m,_\perp}^\nu\,.
\end{align}
The PV reduction reads
\begin{align}
\int [\text{d}l]\frac{X}{\prod_{j}\text{D}_j}\prod_{k'}^n l_{i_{k'},\perp}^{\mu_{k'}}=\begin{cases}
0\,,& n\text{ odd}\\
\sum_\sigma \left[\int [\text{d}l]\frac{X}{\prod_{j}\text{D}_j}\right]y_\sigma(l_\perp\cdot l_\perp,d) \eta_\perp^{\mu_{\sigma_1}\mu_{\sigma_2}}\cdots\eta_\perp^{\mu_{\sigma_{n-1}}\mu_{\sigma_n}}\,,& n\text{ even}
\end{cases}\,.
\label{tensor decomposition}
\end{align}
where the sum of $\sigma$ runs over all the inequivalent permutations of $\mu_1\cdots\mu_n$. The factor $\int [\text{d}l]\frac{X}{\prod_{j}\text{D}_j}$ is just an overall factor. The coefficients $y_\sigma$ can be calculated by contracting both sides of \eqref{tensor decomposition} with $\eta_\perp^{\mu\nu}$ and using the formula $\eta_\perp^{\mu\nu}\eta_{\perp\mu\nu}=d-m$ ($m$ is the number of external momenta). Finally, we substitute
\begin{align}
l_{i,\perp}\cdot l_{j,\perp}=l_i\cdot l_j-\sum_{k,s}^{m}c_{i,k}c_{j,s}\ p_k\cdot p_s\,,
\label{lperpsquare}
\end{align}
for all $l_\perp\cdot l_\perp$ which would appear in $y_\sigma$'s. \eqref{lperpsquare} can be derived by contracting $l_i$ and $l_j$ and then applying \eqref{l decompse}.

In this way, we complete the tensor reduction for \eqref{PVproblem}, and the resulting form is ready for the IBP reduction. In conclusion, the method can be divided into three steps:
\begin{enumerate}
\item 
We first do the momentum decomposition \eqref{l decompse}. 
\item 
Then we do the PV reduction according to \eqref{tensor decomposition}. 
\item 
Finally we substitute \eqref{lperpsquare} for all $l_\perp\cdot l_\perp$.
\end{enumerate}
As a remark, the loop momentum decomposition and the PV reduction are only related to the external momenta appearing in denominators. In other words, the calculation is not sensitive to the number of external legs outside the loops. One can see from Figure~\ref{1loop_cuts} and Figure~\ref{2loop_cuts} that for our calculation, the maximum number of external momenta in denominators is three. Therefore, this method turns out to be efficient for the two-loop calculation of length-4 and 5 operators in our work. The method can be straightforwardly applied to operators of higher lengths.

\section{Anomalous dimensions of the dimension-10 operators}\label{sec:result}

In this section, we present the results of the renormalization matrices and the anomalous dimensions for the dimension-10 operators. 
In Section~\ref{getZ}, we give the results in the $\overline{\text{MS}}$ scheme and the finite renormalization scheme up to the two-loop order. In Section~\ref{sec:fixed point}, we present the anomalous dimensions at the Wilson-Fisher conformal fixed point up to the next-to-leading order. At the fixed point, the anomalous dimensions are independent of the renormalization scheme.
\subsection{The dimension-10 $Z$ matrix and anomalous dimensions }\label{getZ}
Our results include the $Z^{(1)}_{L\to L}$, $Z^{(1)}_{L\to L+1}$, $Z^{(2)}_{L\to L-1}$ and $Z^{(2)}_{L\to L}$ blocks of the $Z$ matrix and the anomalous dimensions up to the two-loop order.\footnote{One will see that the $Z^{(1)}_{2\to 3}$ is not presented in the following. This is because the only length-2 operator is a total derivative of $F^2$, leading to the fact that all $Z_{2\to n}$ vanishes if $n>2$.} We present the results in the $\overline{\text{MS}}$ scheme in Section~\ref{2loopadms}, then the results in the finite renormalization scheme in Section~\ref{2loopad}. One can refer to \eqref{Zstructure}$\sim$\eqref{Zee} for our arrangement of the Z matrix.

\subsubsection{The $\overline{\text{MS}}$ scheme}\label{2loopadms}

In this section, we present the results in the $\overline{\text{MS}}$ scheme. We first review the one-loop $Z$ matrix and the one-loop anomalous dimensions $\gamma^{(1)}$, which were computed in \cite{Jin:2022ivc}.\footnote{The one-loop renormalization for dimension 6 and 8 YM operators were given in \cite{Gracey:2002he,Morozov:1984goy, Neill:2009tn, Harlander:2013oja, Dawson:2014ora}. The two-loop renormalization for dimension-6 operators  were considered in \cite{Jin:2018fak, Jin:2019opr}, and the two-loop renormalization for length-3 operators up to dimension 16 were obtained in \cite{Jin:2020pwh}.} Then we present our result of the two-loop physical anomalous dimensions $\gamma^{(2)}$.

Below is the one-loop result. The blocks in $Z^{\text{even},(1)}_{\text{pp}}$ are
\begin{align}
&Z^{\text{even},(1)}_{\text{pp},2\to 2}=\frac{N_c}{\epsilon}
\left(
\begin{array}{c}
-\frac{11}{3 } \\
\end{array}
\right)\,,
\quad
Z^{\text{even},(1)}_{\text{pp},3\to 3}=\frac{N_c}{\epsilon}
\left(
\begin{array}{cccc}
3 & 0 & 0 & 0 \\
-\frac{3}{5} & \frac{21}{5} & 0 & 0 \\
0 & 0 & \frac{7}{3} & 0 \\
0 & 0 & -1 & \frac{14}{3} \\
\end{array}
\right)\,,\label{z1begin}\\
&Z^{\text{even},(1)}_{\text{pp},3\to 4}=\frac{N_c}{\epsilon}
\left(
\begin{array}{ccccccccccccccc}
0 & 0 & 0 & 0 & 0 & 0 & 0 & 0 & 0 & 0 & 0 & 0 & 0 & 0 & 0 \\
0 & 0 & 0 & 0 & 0 & 0 & 0 & 0 & 0 & 0 & 0 & 0 & 0 & 0 & 0 \\
0 & 0 & 0 & 0 & 0 & 0 & 0 & 0 & 0 & 0 & 0 & 0 & 0 & 0 & 0 \\
-\frac{10}{3} & \frac{2}{3  } & 0 & 0 & 0 & -\frac{53}{72  } & -\frac{43}{96 } & 0 & 0 & \frac{7}{8  } & -\frac{1}{8  } & \frac{1}{12  } & 0 & 0 & 0 \\
\end{array}
\right)\,,\\
&Z^{\text{even},(1)}_{\text{pp},4\to 4}=\frac{N_c}{\epsilon}
\left(
\begin{array}{ccccccccccccccc}
0 & \frac{5}{12} & 0 & 0 & 0 & 0 & 0 & 0 & 0 & 0 & 0 & 0 & 0 & 0 & 0 \\
\frac{16}{3} & \frac{17}{3} & 0 & 0 & 0 & 0 & 0 & 0 & 0 & 0 & 0 & 0 & 0 & 0 & 0 \\
\frac{16}{3} & -\frac{5}{12} & \frac{16}{3} & 0 & 0 & 0 & 0 & 0 & 0 & 0 & 0 & 0 & 0 & 0 & 0 \\
0 & \frac{1}{6} & \frac{2}{3} & 8 & -\frac{2}{3} & 0 & 0 & 0 & 0 & 0 & 0 & 0 & 0 & 0 & 0 \\
\frac{8}{3} & -\frac{5}{12} & \frac{2}{3} & -\frac{10}{3} & \frac{14}{3} & 0 & 0 & 0 & 0 & 0 & 0 & 0 & 0 & 0 & 0 \\
0 & 0 & 0 & 0 & 0 & 5 & -\frac{3}{4} & 0 & 0 & 0 & 0 & 0 & 0 & 0 & 0 \\
0 & 0 & 0 & 0 & 0 & -\frac{5}{3} & \frac{9}{4} & 0 & 0 & 0 & 0 & 0 & 0 & 0 & 0 \\
0 & 0 & 0 & 0 & 0 & -\frac{5}{24} & -\frac{3}{8} & \frac{21}{4} & 0 & 0 & 0 & 0 & 0 & 0 & 0 \\
0 & 0 & 0 & 0 & 0 & \frac{1}{6} & \frac{1}{2} & -1 & 6 & 0 & 0 & 0 & 0 & 0 & 0 \\
0 & 0 & 0 & 0 & 0 & 0 & 0 & 0 & 0 & \frac{1}{2} & \frac{1}{2} & 0 & 0 & 0 & 0 \\
0 & 0 & 0 & 0 & 0 & 0 & 0 & 0 & 0 & 1 & \frac{14}{3} & 0 & 0 & 0 & 0 \\
0 & 0 & 0 & 0 & 0 & 0 & 0 & 0 & 0 & 4 & -\frac{1}{2} & \frac{9}{2} & 0 & 0 & 0 \\
0 & 0 & 0 & 0 & 0 & 0 & 0 & 0 & 0 & -\frac{1}{15} & \frac{2}{5} & \frac{1}{60} & \frac{112}{15} & -\frac{1}{6} & \frac{1}{15} \\
0 & 0 & 0 & 0 & 0 & 0 & 0 & 0 & 0 & -\frac{26}{15} & \frac{3}{20} & -\frac{1}{15} & -\frac{1}{5} & \frac{31}{6} & -\frac{4}{15} \\
0 & 0 & 0 & 0 & 0 & 0 & 0 & 0 & 0 & \frac{64}{15} & \frac{2}{5} & \frac{14}{15} & \frac{14}{5} & -\frac{28}{3} & \frac{67}{30} \\
\end{array}
\right)\,,\label{pp44}\\
&Z^{\text{even},(1)}_{\text{pp},4\to 5}=\frac{N_c}{\epsilon}
\left(
\begin{array}{cccc}
0 & 0 & 0 & 0 \\
0 & 0 & 0 & 0 \\
0 & 0 & 0 & 0 \\
-\frac{17}{3} & 8 & 0 & 0 \\
-\frac{26}{3} & 10 & 0 & 0 \\
0 & 0 & 0 & 0 \\
0 & 0 & 0 & 0 \\
0 & 0 & 0 & 0 \\
-\frac{46}{9} & \frac{22}{3} & \frac{2}{3} & \frac{4}{9} \\
0 & 0 & 0 & 0 \\
0 & 0 & 0 & 0 \\
0 & 0 & 0 & 0 \\
0 & 0 & -\frac{1}{5} & -\frac{1}{15} \\
0 & 0 & -\frac{6}{5} & -\frac{7}{30} \\
0 & 0 & -\frac{197}{10} & -\frac{169}{60} \\
\end{array}
\right)\,,
\quad
Z^{\text{even},(1)}_{\text{pp},5\to 5}=\frac{N_c}{\epsilon}
\left(
\begin{array}{cccc}
-\frac{11}{3} & 10 & 0 & 0 \\
-9 & \frac{49}{3} & 0 & 0 \\
0 & 0 & \frac{9}{2} & \frac{1}{4} \\
0 & 0 & 1 & \frac{37}{6} \\
\end{array}
\right)\,.
\end{align}
One can see that at the one-loop order, there is no mixing between different helicity sectors.

The blocks in $Z_{\text{pp}}^{\text{odd},(1)}$ are
\begin{align}
&Z^\text{odd,(1)}_{\text{pp},3\to 3}=\frac{N_c}{\epsilon}
\left(
\begin{array}{c}
4 \\
\end{array}
\right)\,,
Z^\text{odd,(1)}_{\text{pp},3\to 4}=\frac{N_c}{\epsilon}
\left(
\begin{array}{ccccc}
0 & 0 & 0 & 0 & 0 \\
\end{array}
\right)\,,
Z^\text{odd,(1)}_{\text{pp},4\to 4}=\frac{N_c}{\epsilon}
\left(
\begin{array}{ccccc}
\frac{16}{3} & 0 & 0 & 0 & 0 \\
0 & \frac{17}{4} & 0 & 0 & 0 \\
0 & -1 & \frac{25}{4} & 0 & 0 \\
0 & 0 & 0 & \frac{37}{10} & -\frac{1}{5} \\
0 & 0 & 0 & -\frac{3}{10} & \frac{82}{15} \\
\end{array}
\right)\,.
\end{align}

The blocks in $Z_{\text{pe}}^{\text{even},(1)}$ are
\begin{align}
&Z^{\text{even},(1)}_{\text{pe},3\to 4}=\frac{N_c}{\epsilon}
\left(
\begin{array}{ccc}
0 & 0 & 0 \\
0 & 0 & 0 \\
0 & 0 & 0 \\
\frac{17}{48 } & -\frac{5}{192 } & 0 \\
\end{array}
\right)\,,
&&Z^{\text{even},(1)}_{\text{pe},4\to 4}=\frac{N_c}{\epsilon}
\left(
\begin{array}{ccc}
0 & 0 & 0 \\
0 & 0 & 0 \\
-\frac{1}{3} & -\frac{1}{4} & 0 \\
-\frac{41}{72} & -\frac{23}{36} & -\frac{7}{12} \\
\frac{19}{72} & \frac{13}{36} & \frac{5}{12} \\
-\frac{1}{18} & \frac{13}{72} & 0 \\
-\frac{1}{54} & \frac{49}{216} & 0 \\
-\frac{143}{864} & -\frac{41}{432} & 0 \\
\frac{43}{216} & \frac{7}{27} & \frac{1}{4} \\
0 & 0 & 0 \\
0 & 0 & 0 \\
-\frac{19}{18} & -\frac{29}{72} & 0 \\
-\frac{1}{240} & -\frac{1}{80} & -\frac{1}{60} \\
-\frac{1}{240} & \frac{1}{20} & \frac{1}{40} \\
-\frac{27}{80} & -\frac{11}{80} & -\frac{19}{40} \\
\end{array}
\right)\,,\\
&Z^{\text{even},(1)}_{\text{pe},4\to 5}=\frac{N_c}{\epsilon}
\left(
\begin{array}{cc}
0 & 0 \\
0 & 0 \\
0 & 0 \\
\frac{65}{18} & \frac{31}{18} \\
-\frac{11}{18} & -\frac{23}{9} \\
0 & 0 \\
0 & 0 \\
0 & 0 \\
\frac{10}{27} & -\frac{46}{27} \\
0 & 0 \\
0 & 0 \\
0 & 0 \\
-\frac{11}{20} & \frac{1}{10} \\
\frac{281}{180} & -\frac{1}{90} \\
\frac{277}{60} & -\frac{63}{20} \\
\end{array}
\right)\,,
&&Z^{\text{even},(1)}_{\text{pe},5\to 5}=\frac{N_c}{\epsilon}
\left(
\begin{array}{cc}
\frac{25}{18} & -\frac{55}{18} \\
\frac{10}{9} & -\frac{19}{9} \\
\frac{1}{6} & -\frac{5}{6} \\
-\frac{2}{9} & -\frac{4}{9} \\
\end{array}
\right)\,.
\end{align}

The blocks in $Z_{\text{pe}}^{\text{odd},(1)}$ are
\begin{align}
Z^\text{odd,(1)}_{\text{pe},3\to 4}=\frac{N_c}{\epsilon}
\left(
\begin{array}{c}
0 \\
\end{array}
\right)\,,
Z^\text{odd,(1)}_{\text{pe},4\to 4}=\frac{N_c}{\epsilon}
\left(
\begin{array}{c}
\frac{1}{12} \\
0 \\
-\frac{1}{16} \\
\frac{19}{120} \\
\frac{13}{40} \\
\end{array}
\right)\,.
\end{align}

The blocks in $Z_{\text{ee}}^{\text{even},(1)}$ are
\begin{align}
Z^{\text{even},(1)}_{\text{ee},4\to 4}=\frac{N_c}{\epsilon}
\left(
\begin{array}{ccc}
3 & -\frac{8}{3} & 0 \\
-\frac{10}{3} & \frac{14}{3} & 0 \\
2 & 1 & \frac{19}{3} \\
\end{array}
\right)\,,
Z^{\text{even},(1)}_{\text{ee},4\to 5}=\frac{N_c}{\epsilon}
\left(
\begin{array}{cc}
0 & 0 \\
0 & 0 \\
-\frac{112}{9} & -\frac{20}{9} \\
\end{array}
\right)\,,
Z^{\text{even},(1)}_{\text{ee},5\to 5}=\frac{N_c}{\epsilon}
\left(
\begin{array}{cc}
\frac{32}{9} & -\frac{26}{9} \\
-\frac{55}{9} & \frac{52}{9} \\
\end{array}
\right)\,.
\end{align}

The only block in $Z_{\text{ee}}^{\text{odd},(1)}$ is
\begin{align}
Z^\text{odd,(1)}_{\text{ee},5\to 5}=\frac{N_c}{\epsilon}
\left(
\begin{array}{c}
5 \\
\end{array}
\right).
\label{zee5}
\end{align}

As discussed in Section~\ref{zms}, the block $Z^{(1)}_{\text{ep}}$ is zero in the $\overline{\text{MS}}$ scheme. This can also be understood by the fact that the one-loop four-dimensional cut of an evanescent operator vanishes. In each $Z_{L\to L}$ block, the $Z$-matrix turns out to be block upper triangular within each helicity sector. This is due to the fact that we enumerate all the possible total derivative operators in our operator basis. One can refer to Appendix~\ref{all dim-10} for a more detailed discussion about total derivative operators. 

According to \eqref{gamma1}, one gets $\mathcal{D}^{(1)}$. Since $\mathcal{D}^{(1)}_{\text{ep}}$ vanishes, the physical and evanescent anomalous dimensions are just eigenvalues of $\mathcal{D}^{(1)}_{\text{pp}}$ and $\mathcal{D}^{(1)}_{\text{ee}}$ respectively. Within each block, there is no mixing between the C-even and C-odd operators. Besides, each C-parity block is upper triangular according to the length. Therefore, the anomalous dimensions can be further classified according to C-parities and lengths. The one-loop anomalous dimensions are given as follows, where each anomalous dimension includes an implicit factor $N_c$:
\begin{align}
&\gamma_{\text{p,length-2}}^{\text{even},(1)}:-\frac{22}{3}\,,\label{p2eveng1}\\
&\gamma_{\text{p,length-3}}^{\text{even},(1)}:\frac{14}{3},6,\frac{42}{5},\frac{28}{3}\,,\\
&\gamma_{\text{p,length-3}}^{\text{odd},(1)}:\  8\,,\\
&\gamma_{\text{p,length-4}}^{\text{even},(1)}: 9,\frac{21}{2},\frac{32}{3},12,\frac{1}{3} \left(17\pm 3 \sqrt{41}\right),\frac{1}{6} \left(31\pm \sqrt{697}\right),\frac{1}{4} \left(29\pm \sqrt{201}\right),\frac{2}{3} \left(19\pm 3 \sqrt{5}\right),\nonumber\\
&\qquad\qquad\qquad\quad
x_1, x_2,x_3\,,
\\
&\gamma_{\text{p,length-4}}^{\text{odd},(1)}:\  \frac{22}{3},\frac{17}{2},\frac{32}{3},11,\frac{25}{2}\,,\\
&\gamma_{\text{p,length-5}}^{\text{even},(1)}:\frac{2}{3} \left(19\pm 3 \sqrt{10}\right),\frac{1}{3} \left(32\pm \sqrt{34}\right)\,,\\
&\gamma_{\text{e,length-4}}^{\text{even},(1)}=\frac{1}{3}\left(23\pm\sqrt{345}\right),\frac{38}{3}\,,\\
&\gamma_{\text{e,length-5}}^{\text{even},(1)}=\frac{2}{3} \left(14\pm\sqrt{170}\right)\,,\\
&\gamma_{\text{e,length-4}}^{\text{odd},(1)}=10\,,\label{e4oddg1}
\end{align}
where the subscript ``p"(``e") means ``physical"(``evanescent"). The $x_i$ are the solutions of the equation
\begin{align}
x^3-\frac{446 x^2}{15}+\frac{769 x}{3}-\frac{8014}{15}=0\,,
\end{align}
with $x_1<x_2<x_3$. Their numerical solutions are
\begin{align}
x_1=3.0565 \,,\quad\  x_2=11.573 \,,\quad\  x_3=15.104 \,.
\end{align}

At the two-loop order, the $Z$ matrix is calculated as shown in Section~\ref{zms}. The dilatation matrix can then be calculated according to \eqref{gamma2}. The results are given in the auxiliary file. As an example, we also present the block $Z^{(2)}_{\text{pp}}$ in Appendix~\ref{zppresult}. Below are the 2-loop corrections of the ones shown in \eqref{p2eveng1}$\sim$\eqref{e4oddg1}, where each anomalous dimension includes an implicit factor $N_c^2$:
\begin{align}
&\gamma_{\text{p,length-2}}^{\text{even},(2)}:-\frac{136}{3}\,,\label{l2eveng1}\\
&\gamma_{\text{p,length-3}}^{\text{even},(2)}:\frac{59}{3},\frac{439}{18},\frac{7121}{250},\frac{149525}{3996}\,,\\
&\gamma_{\text{p,length-3}}^{\text{odd},(2)}:\  \frac{206}{9}\,,\label{l3MSodd}\\
&\gamma_{\text{p,length-4}}^{\text{even},(2)}: \frac{1308521}{35532},\frac{12319}{288},\frac{815}{18},\frac{415}{18},\frac{37679\pm 2651 \sqrt{41}}{1476},\frac{2 \left(179129\pm 2352 \sqrt{697} \right)}{18819},\nonumber\\
&\qquad\qquad\qquad\quad
\frac{29 \left(4108755061 \sqrt{201}\pm 115875887553\right)}{112160431296},\frac{1}{54} \left(3100\pm 103 \sqrt{5}\right),y_1, y_2,y_3\,,
\\
&\gamma_{\text{p,length-4}}^{\text{odd},(2)}:\  \frac{32885}{1188},\frac{3125}{96},\frac{107}{2},\frac{75421}{1188},\frac{64211}{1440}\,,\\
&\gamma_{\text{p,length-5}}^{\text{even},(2)}:\frac{376249\pm 78535 \sqrt{10}}{8604},\frac{108341113246123 \sqrt{34}\pm 4211644375821510}{113297323414176}\,,\\
\label{l5eveng1}
&\gamma_{\text{e,length-4}}^{\text{even},(2)}:\frac{\left(6442724032485\pm11542242689 \sqrt{345}\right)}{213976901880},\frac{4755559}{75255}\,,\\
&\gamma_{\text{e,length-5}}^{\text{even},(2)}:\frac{\left(3977690861205\pm50021112896 \sqrt{170}\right)}{114158809836}\,,\\
&\gamma_{\text{e,length-4}}^{\text{odd},(2)}:\frac{3079}{540}\,.\label{e4oddg2}
\end{align}
The $y_i$ are the solutions of the equation
\begin{align}
y^3-\frac{44053970579731 y^2}{334691552250}&+\frac{4335623758063848120847262203 y}{800852671362744392040000}\nonumber\\
&-\frac{12858742227506943574716057437659}{194607199141146887265720000}=0\,,\label{yms}
\end{align}
with $y_1<y_2<y_3$. Their numerical solutions are
\begin{align}
y_1=22.029\,,\quad\ y_2=52.952\,,\quad\ y_3=56.644\,.
\end{align}

An observation is that almost all the anomalous dimensions are positive. There are only two exceptions. One of them is the anomalous dimension of $\text{tr}(F^2)$, which can be written through the $\beta$ function~\cite{Spiridonov:1988md} and is negative at the one-loop and the two-loop order. The other one is $\frac{1}{3} \left(17-3 \sqrt{41}\right)$, which is one of the $\gamma_{\text{p,length-4}}^{\text{even},(1)}$. (While its two-loop correction is positive.) It would be interesting to understand better the signs of anomalous dimensions.

\subsubsection{The finite renormalization scheme}\label{2loopad}

In this section, we present the results in the finite renormalization scheme. Following Section~\ref{zfin}, we use $\hat{Z}$ to denote the $Z$ matrix and $\hat{\gamma}$ to denote the anomalous dimensions in the finite renormalization scheme. 

The one-loop $Z$ matrix includes four blocks
\begin{align}
	\left(
	\begin{array}{cc}
		\hat{Z}_{\text{pp}}^{(1)} & \hat{Z}_{\text{pe}}^{(1)} \\
		\hat{Z}_{\text{ep}}^{(1)} & \hat{Z}_{\text{ee}}^{(1)}
	\end{array}
	\right)\,.
\end{align} 
The blocks $\hat{Z}_{\text{pp}}^{(1)}$, $\hat{Z}_{\text{pe}}^{(1)}$ and $\hat{Z}_{\text{ee}}^{(1)}$ are the same as the ones in the $\overline{\text{MS}}$ scheme. The only difference is the block $\hat{Z}_{\text{ep}}^{(1)}$, which is finite in this scheme. The blocks in $\hat{Z}^{\text{even},(1)}_{\text{ep}}$ read
\begin{align}
&\hat{Z}^{\text{even},(1)}_{\text{ep},4\to 4}=N_c
\left(
\begin{array}{ccccccccccccccc}
\frac{16}{3} & -\frac{2}{3} & 0 & 0 & 0 & \frac{10}{3} & -2 & 0 & 0 & \frac{16}{3} & -\frac{14}{3} & \frac{14}{3} & 0 & 0 & 0 \\
0 & 1 & 0 & 0 & 0 & -\frac{1}{3} & 0 & 0 & 0 & -\frac{2}{3} & \frac{7}{3} & -\frac{1}{3} & 0 & 0 & 0 \\
0 & -\frac{2}{3} & 0 & -\frac{4}{3} & \frac{4}{3} & \frac{2}{9} & -\frac{3}{2} & \frac{10}{3} & -\frac{14}{3} & \frac{26}{9} & 0 & \frac{1}{9} & \frac{28}{3} & -\frac{58}{9} & -\frac{14}{9} \\
\end{array}
\right)\,,\\
&\hat{Z}^{\text{even},(1)}_{\text{ep},4\to 5}=N_c
\left(
\begin{array}{cccc}
0 & 0 & 0 & 0 \\
0 & 0 & 0 & 0 \\
-\frac{182}{9} & \frac{92}{3} & -\frac{50}{3} & -4 \\
\end{array}
\right)\,,
\hat{Z}^{\text{even},(1)}_{\text{ep},5\to 5}=N_c
\left(
\begin{array}{cccc}
\frac{14}{3} & -\frac{17}{3} & \frac{10}{3} & -\frac{1}{2} \\
-\frac{35}{6} & \frac{25}{3} & -\frac{5}{3} & \frac{5}{4} \\
\end{array}
\right)\,.
\end{align}

The only block in $\hat{Z}^{\text{odd},(1)}_{\text{ep}}$ read
\begin{align}
\hat{Z}^{\text{odd},(1)}_{\text{ep},4\to 4}=N_c
\left(
\begin{array}{ccccc}
-\frac{8}{3} & 0 & -3 & \frac{4}{3} & -\frac{14}{3} \\
\end{array}
\right)\,.
\end{align}

Since the difference between $\hat{Z}^{(1)}$ and ${Z}^{(1)}$ is finite, the difference between the corresponding dilatation matrices is of order $\epsilon$ according to \eqref{gamma1}. Therefore, the one-loop anomalous dimensions are the same in the two schemes. The anomalous dimensions are the same as the ones in \eqref{p2eveng1}$\sim$\eqref{e4oddg1}.

The two-loop $Z$ matrix can be calculated as described in Section~\ref{zfin}. The blocks $\hat{Z}^{(2)}_{\text{pp}}$ and $Z^{(2)}_{\text{pe}}$ are the same as the ones in the $\overline{\text{MS}}$ scheme, while $Z^{(2)}_{\text{ep}}$ and $Z^{(2)}_{\text{ee}}$ are different due to the contribution of the finite $Z^{(1)}_{\text{ep}}$. In this scheme, the dilatation matrix is block upper triangular as shown in \eqref{uptriangulargamma}. As discussed in Section~\ref{zfin}, this does not mean the evanescent operators are irrelevant to the physical anomalous dimensions, since $\hat{\mathcal{D}}^{(2)}_{\text{pp}}$ receives the contribution from the term $(-2\epsilon {\hat{Z}}^{(1)}_{\text{pe}}{\hat{Z}}^{(1)}_{\text{ep}})$. Note that if one only need to calculate the physical anomalous dimensions, only $\hat{Z}_{\text{pp}}^{(2)}$ is required, whose expression can be found in Appendix~\ref{zppresult}. Other blocks of the two-loop $Z$ matrix are given in the auxiliary file.

Below are the two-loop anomalous dimensions, where each anomalous dimension includes an implicit factor $N_c^2$:
\begin{align}
&\hat{\gamma}_{\text{p,length-2}}^{\text{even},(2)}:-\frac{136}{3}\,,\label{l2finite}\\
&\hat{\gamma}_{\text{p,length-3}}^{\text{even},(2)}:\frac{59}{3},\frac{439}{18},\frac{7121}{250},\frac{149525}{3996}\,,\label{l3finiteeven}\\
&\hat{\gamma}_{\text{p,length-3}}^{\text{odd},(2)}: \frac{206}{9}\,,\label{l3finiteodd}\\
&\hat{\gamma}_{\text{p,length-4}}^{\text{even},(2)}:\frac{3427}{108},\frac{12319}{288},\frac{815}{18},\frac{877}{18},\frac{37679\pm 2651 \sqrt{41}}{1476},\frac{2 \left(179129\pm 2352 \sqrt{697}\right)}{18819},\nonumber\\
&\qquad\qquad\qquad\quad
\frac{129729219\pm 5049167 \sqrt{201}}{4283712},\frac{1}{270} \left(16160\pm 251 \sqrt{5}\right),y_1,y_2,y_3\,,\\
&\hat{\gamma}_{\text{p,length-4}}^{\text{odd},(2)}: \frac{8048}{297},\frac{3125}{96},\frac{875}{18},\frac{50825}{1188},\frac{13159}{288}\,,\\
&\hat{\gamma}_{\text{p,length-5}}^{\text{even},(2)}:\frac{1}{18} \left(617\pm 142 \sqrt{10}\right),\frac{432664955274\pm 20543721361 \sqrt{34}}{12629285856}\,,\label{l5finite}\\
&\hat{\gamma}_{\text{e,length-4}}^{\text{even},(2)}: \frac{97}{3}\mp\frac{59 \sqrt{\frac{5}{69}}}{3},\frac{9098 }{261}\,,\\
&\hat{\gamma}_{\text{e,length-5}}^{\text{even},(2)}: \frac{\left(2513637\pm54631 \sqrt{170}\right)}{53244}\,,\\
&\hat{\gamma}_{\text{e,length-4}}^{\text{odd},(2)}: \frac{277}{9}\,.
\end{align}
The $y_i$ are the solutions of the equation
\begin{align}
y^3-\frac{250350031847 y^2}{1934633250}&+\frac{4824966722800230692858971 y}{925841238569646696000}\nonumber\\
&-\frac{13849613580264790328390513509}{224979420972424147128000}=0\,,\label{yfinite}
\end{align}
with $y_1<y_2<y_3$. Their numerical solutions are
\begin{align}
y_1=20.933\,,\quad\ y_2=53.407\,,\quad\ y_3=55.065\,.
\end{align}

As in the $\overline{\text{MS}}$ scheme, the two-loop anomalous dimensions are all positive in the finite scheme except for $\gamma_{\text{p,length-2}}^{\text{even},(2)}$. From the above results, one can see that all the two-loop \emph{evanescent} anomalous dimensions are different from the ones in the $\overline{\text{MS}}$ scheme. While due to the relations $\hat{Z}_{\text{pp}}=Z_{\text{pp}}$ and $\hat{Z}_{\text{pe}}=Z_{\text{pe}}$, some physical anomalous dimensions remain the same in the two schemes. From \eqref{l2eveng1}$\sim$\eqref{l3MSodd} and \eqref{l2finite}$\sim$\eqref{l3finiteodd}, we see that all length-2 and length-3 two-loop anomalous dimensions are the same in the two schemes. The reason is that there is no mixing from evanescent operators to length-2 and length-3 operators up to the two-loop order.\footnote{Here the mixing means divergent mixing. Actually, there can be finite mixing from an evanescent operator to length-3 operators in the finite renormalization scheme. However, this mixing would not affect the two-loop anomalous dimensions.} Besides, some length-4 anomalous dimensions remain the same in the two schemes. This is due to the fact that the $Z$ matrix is block upper triangular according to the $D$-type (a detailed discussion is given in Appendix~\ref{all dim-10}) and these length-4 operators are in the $D$-type sectors where there is no evanescent operator.

\subsection{Anomalous dimensions at the conformal fixed point}\label{sec:fixed point}

From \eqref{l2eveng1}$\sim$\eqref{yms} and \eqref{l2finite}$\sim$\eqref{yfinite} one can see that the two-loop anomalous dimensions depend on the renormalization scheme. On the other hand, the anomalous dimensions at a conformal fixed point should not depend on the scheme choice (see \emph{e.g.} \cite{Vasilev:2004yr} and \cite{DiPietro:2017vsp}). This provides a non-trivial crosscheck between our results in the two schemes. In the following, we first give a short proof for why the anomalous dimension is scheme independent at a conformal fixed point. Then we show that given $\mathcal{D}^{(1)}$ and $\mathcal{D}^{(2)}$, how to calculate the dilatation matrix up to next-to-leading (NLO) at the WF fixed point. 
Finally, we give all anomalous dimensions of dimension-10 operators at the WF fixed point. We use $\mathcal{D}^*$ and $\gamma^*$ to denote the dilatation matrix and anomalous dimensions at the WF fixed point. 

Assume that we have a set of renormalized operators,
\begin{align}
	O_j=({Z})_j^{\ k}O_{k,\text{b}}\,.
\end{align}
A change of subtraction scheme can be generally thought of as a finite linear transformation of them \cite{Vasilev:2004yr}:
\begin{align}
	{K}_i^{\ j}O_j={K}_i^{\ j}({Z})_j^{\ k}O_{k,\text{b}}\,,
	\label{lineartransofO}
\end{align}
Define $\tilde{Z}\equiv{K}{Z}$ and together with \eqref{overallgamma}, one can get $\tilde{\mathcal{D}}$ as
\begin{align}
	\tilde{\mathcal{D}}=-\frac{\partial{K}}{\partial \alpha_s}(\mu\frac{d\alpha_s}{d{\mu}}){K}^{-1}+{K}\mathcal{D}{K}^{-1}\,,
	\label{gammalineartrans}
\end{align}
with $\mathcal{D}=-\mu\frac{\text{d}Z}{\text{d}\mu}(Z)^{-1}$. The transformation is not a similarity transformation in general, so the eigenvalues of $\mathcal{D}$ and the ones of $\tilde{\mathcal{D}}$ are different. But if the theory is at a conformal fixed point, one has $\mu\frac{d\alpha_s}{d{\mu}}=0$. Then a scheme transformation leads to a similar transformation of the dilatation matrix, leading to the fact that the anomalous dimensions, are independent of the renormalization scheme.

Below we show the calculation of the anomalous dimensions at the WF fixed point. According to \eqref{renormalpha}, the coupling at the WF fixed point reads
\begin{align}
\alpha^*=-\frac{4 \pi  \epsilon }{\beta_0}-\frac{4 \pi  \beta_1 \epsilon ^2}{\beta_0^3}+\mathcal{O}(\epsilon^3)\,.
\label{alphaWF}
\end{align}
{We would like to point out that the pure YM theory is asympotically free and $\beta_0>0$ as given in \eqref{eq:beta0beta1}, thus the WF fixed point is expected to be a UV fixed point at $d>4$, namely $\epsilon<0$. 
For example, it has been speculated in  \cite{Morris:2004mg,DeCesare:2021pfb} to continue to an interacting CFT in $d = 5$ along this line.}

Substitute the $\mathcal{D}^{(1)}$ and the $\mathcal{D}^{(2)}$ calculated in the last section back into \eqref{overallgamma} and replace $\alpha_s$ by $\alpha^*$, then we get the dilatation matrix expanded in $\epsilon$:
\begin{align}
\mathcal{D}^*=\sum_{i=1}\epsilon^{i}{\mathcal{D}^*_i}=\sum_{l}\left(\frac{\alpha^*}{4\pi}\right)^l\mathcal{D}^{(l)}=(-\frac{  \epsilon }{\beta_0}-\frac{  \beta_1 \epsilon ^2}{\beta_0^3})\mathcal{D}^{(1)}+\frac{\epsilon^2}{\beta_0^2}\mathcal{D}^{(2)}+\mathcal{O}(\epsilon^3)\,.  \label{Dfix}
\end{align}
In the planar limit, the factor $N_c^l$ in $\mathcal{D}^{(l)}$ cancels the factor $N_c^{-l}$ in $(\alpha^*)^l$, so the dilatation matrix are independent of $N_c$.\footnote{This is a general feature for the anomalous dimensions at the WF fixed point in the large $N_c$ limit, see \emph{e.g.} Chapter 29 in \cite{Zinn-Justin:2002ecy} for the $O(N)$ theory.}
The anomalous dimensions can be expanded in $\epsilon$ as
\begin{align}
\gamma^*=\sum_{i=1}\epsilon^{i}{\gamma^*_i}\,.
\label{gammafix}
\end{align}

According to \eqref{Dfix} and \eqref{gammafix}, one can calculate the dilatation matrix and anomalous dimensions at the fixed point in the $\overline{\text{MS}}$ scheme and the finite renormalization scheme. The dilatation matrices can be found in the auxiliary file. 
{It is worthwhile noting that when considering the WF fixed point, the spacetime dimension $d$ is not equal 4 and one should not take the $\epsilon\to 0$ limit. Therefore when calculating in the finite scheme, the $\hat{\mathcal{D}}^{(l)}_{\text{ep}}$'s which are $\mathcal{O}(\epsilon)$ should be taken into account.}
Our calculation verify that anomalous dimensions in the two schemes are the same. Below we present the anomalous dimensions. 

The leading-order (LO) results are (Note that there is no $N_c^{l}$ factor in $\gamma^*_l$)
\begin{align}
&{\gamma^*_1}^{\text{even}}_{,\text{length-2}}:2\,,\label{fixeven2}\\
&{\gamma^*_1}_{,\text{length-3}}^{\text{even}}:-\frac{28}{11},-\frac{126}{55},-\frac{18}{11},-\frac{14}{11}\,,\\
&{\gamma^*_1}_{,\text{length-3}}^{\text{odd}}:\ -\frac{24}{11} \,,\\
&{\gamma^*_1}_{,\text{length-4}}^{\text{even}}: -\frac{38}{11},-\frac{36}{11},-\frac{32}{11},-\frac{63}{22},-\frac{27}{11},
\frac{2}{11} \left(-19\pm 3 \sqrt{5}\right),\frac{1}{11} \left(-17\pm 3 \sqrt{41}\right),\nonumber\\
&\qquad\qquad\qquad\quad
\frac{3}{44} \left(-29\pm \sqrt{201}\right),\frac{1}{11} \left(-23\pm\sqrt{345}\right),\frac{1}{22} \left(-31\pm\sqrt{697}\right),x_1,x_2,x_3\,,
\\
&{\gamma^*_1}_{,\text{length-4}}^{\text{odd}}:\ -\frac{75}{22},-3,-\frac{32}{11},-\frac{30}{11},-\frac{51}{22},-2\,, \\
&{\gamma^*_1}_{,\text{length-5}}^{\text{even}}:\frac{2}{11} \left(-19\pm 3 \sqrt{10}\right),\frac{2}{11} \left(-14\pm\sqrt{170}\right),\frac{1}{11} \left(-32\pm \sqrt{34}\right)\,,\label{fixeven5}
\end{align}
The $x_i$ are the roots of
\begin{align}
x^3+\frac{446 x^2}{55}+\frac{2307 x}{121}+\frac{72126}{6655}=0\,,
\end{align}
with $x_1<x_2<x_3$. Their numerical solutions are
\begin{align}
x_1=-4.1193\,,\quad\ x_2=-3.1562\,,\quad\ x_3=-0.83360\,.
\end{align}
Actually, one can derive from \eqref{Dfix} that the LO results can be achieved by substituting $\frac{\alpha_s N_c}{4\pi}\to -\frac{3}{11}$ into the one-loop anomalous dimensions calculated in the $\alpha_s$-expansion.

The next-to-leading (NLO) corrections are
\begin{align}
&{\gamma^*_2}_{,\text{length-2}}^{\text{even}}:-\frac{204}{121}\,,\\
&{\gamma^*_2}_{,\text{length-3}}^{\text{even}}:\frac{376711}{590964},\frac{62379}{332750},\frac{1157}{2662},\frac{519}{1331} \,,\\
&{\gamma^*_2}_{,\text{length-3}}^{\text{odd}}:\ -\frac{182}{1331} \,,\\
&{\gamma^*_2}_{,\text{length-4}}^{\text{even}}: \frac{59703987}{33388135},-\frac{2779}{2662},\frac{2437}{2662},\frac{32693}{42592},\frac{3520939}{5254788},\frac{10844\pm 4805 \sqrt{5}}{7986}\nonumber\\
&\qquad\qquad\qquad\quad
\frac{130093\pm 21023 \sqrt{41}}{218284},\frac{9316861814943\pm 357329198443 \sqrt{201}}{16587281561664},\nonumber\\
&\qquad\qquad\qquad\quad
\frac{15093318600615\pm 2298106885061 \sqrt{345}}{31644806266920},
\frac{634967\pm 54897 \sqrt{697}}{2783121}
\nonumber\\
&\qquad\qquad\qquad\quad
,y_1,y_2,y_3\,,
\\
&{\gamma^*_2}_{,\text{length-4}}^{\text{odd}}:\ \frac{94321}{212960},\frac{35029}{15972},\frac{4065}{2662},-\frac{149731}{79860},\frac{19893}{42592},\frac{5957}{15972}\,, \\
&{\gamma^*_2}_{,\text{length-5}}^{\text{even}}:\frac{278813 \sqrt{10}\pm 433283}{1272436}, \frac{7528203818631\pm 2037367447760 \sqrt{170}}{16882819543524},\nonumber\\
&\qquad\qquad\qquad\quad
\frac{476265955378374\pm 8389462392725 \sqrt{34}}{1523219570346144}\,,
\end{align}
The $y_i$ are the roots of
\begin{align}
y^3-\frac{13294802711131 y^2}{4499741980250}&+\frac{4515566752618635673017361833 y}{1592322513279522803486840000}\nonumber\\
&-\frac{10138741527932683336204035731219}{11444658831945242197781313816000}=0\,,
\end{align}
with $y_1<y_3<y_2$. Their numerical solutions are
\begin{align}
y_1=0.740685\,,\quad\ y_2=1.27805\,,\quad \ y_3=0.935839\,.
\end{align}

From \eqref{Dfix}, one can see that the $\mathcal{D}^*_2$ includes the one-loop term $-\frac{\beta_1}{\beta_0^3} \mathcal{D}^{(1)}$.\footnote{In the finite renormalization, the term $-\frac{1}{\beta_0}\hat{\mathcal{D}}_{\text{ep}}^{(1)}$ would also contribute to $\mathcal{D}^*_2$.}
This may alter the signs of the NLO anomalous dimensions. Let us take the length-3 C-odd operator $O_{25}$ as an example. Since $O_{25}$ is an eigenstate of the dilatation matrix (as presented in Appendix~\ref{all dim-10}), in this simple case one can get its anomalous dimension at the fixed point via the following formula
\begin{align}
{\gamma^*_2}_{,\text{length-3}}^{\text{odd}}&=(\gamma_{\text{p},\text{length-3}}^{\text{odd},(1)}\alpha^*+\gamma_{\text{p},\text{length-3}}^{\text{odd},(2)}{\alpha^*}^2)\big{|}_{\text{coefficients of }\epsilon^2}\nonumber\\
&=\gamma_{\text{p},\text{length-3}}^{\text{odd},(1)}(-\frac{\beta_1}{\beta_0^3})+\gamma_{\text{p},\text{length-3}}^{\text{odd},(2)}\frac{1}{\beta_0^2}\nonumber\\
&=8(-\frac{306}{1331})+\frac{206}{9}\frac{9}{121}=-\frac{182}{1331}\,.
\end{align}
One can see that the term $\gamma_{\text{p},\text{length-3}}^{\text{odd},(1)}(-\frac{\beta_1}{\beta_0^3})$ alter the sign of the NLO anomalous dimension.

\section{Conclusion}\label{sec:discuss}

In this paper, we study the two-loop renormalization of gluonic evanescent operators in the pure YM theory. Although the tree-level matrix elements of evanescent operators vanish in four-dimensional spacetime, they are important at the quantum loop level in dimensional regularization since the internal legs can propagate in $d=4-2\epsilon$ dimensions. The effect of the evanescent operators on the physical anomalous dimensions comes from their mixing with physical ones, while the pattern of the effect depends on the renormalization scheme. Let us take the $\overline{\text{MS}}$ scheme as an example. At the one-loop order, such mixing is suppressed to be finite by the evanescent effect. Thus the evanescent operators have no effect on the physical anomalous dimensions.
At the two-loop order, however, the mixing can be of order $1/\epsilon$ and can give rise to an important contribution to the physical anomalous dimensions.
Our two-loop computation for the dimension-10 operator basis provides a first concrete example in the Yang-Mills theory to show the effect of gluonic evanescent operators on the physical anomalous dimensions.

We have applied two different schemes to obtain the anomalous dimensions. 
In the $\overline{\text{MS}}$ scheme, one needs to consider the full renormalization matrix of both physical and evanescent operators up to the two-loop order.
In the finite renormalization scheme, the dilatation matrix has the nice property that it is block upper triangular \cite{Buras:1989xd,Dugan:1990df};
therefore, to compute physical anomalous dimensions, one only needs to consider physical operators up to the two-loop order. However, we stress that it is still necessary to compute the one-loop renormalization of evanescent operators. More generally, to compute $l$-loop physical anomalous dimensions, one needs to consider the renormalization of evanescent operators up to the $(l-1)$-loop orders in the finite renormalization scheme.
The anomalous dimensions depend on the renormalization scheme due to the running effect of the coupling constant.
As a further consideration, we compute the anomalous dimensions for the YM theory at the WF fixed point. In this case, the theory is in non-integer dimensions and physical and evanescent operators are on an equal footing. We obtain the anomalous dimensions up to the next-to-leading order in the $\epsilon$-expansion. As expected, we find the anomalous dimensions computed with both renormalization schemes give the same results at the WF fixed point. This also provides a non-trivial check of the results.

For the two-loop computation, we consider form factors which are matrix elements each involving one physical or evanescent operator. 
We use the $d$-dimensional unitarity-cut method combined with efficient integral reduction methods. 
To simplify the computation, we perform numerical computations in the intermediate steps to get numerical UV data and finally reconstruct the analytic $Z$ matrix. 
This provides a first two-loop computation of anomalous dimensions for a close set of Yang-Mills operators which include length-4 and length-5 operators. 
Our strategy can be straightforwardly applied to  the two-loop renormalization of YM operators of higher lengths and is also expected to be applicable for high-dimensional operators in more general theories.

\acknowledgments
This work is supported in part by the National Natural Science Foundation of China (Grants No.~11935013, 12175291, 11822508, 12047503, 12047502, 11947301).
We also thank the support of the HPC Cluster of ITP-CAS.

\begin{appendix}

\section{Dimension-10 operators}\label{all dim-10}

In this appendix, we present our dimension-10 single-trace operator basis. Different from those given in~\cite{Jin:2022ivc}, we organize the basis by separating total derivative operators explicitly.

We first give a complementary discussion about total derivative operators. We say that an operator is of $D$-$(i,\alpha)$ type, if it is an $i$th total derivative of a rank-$\alpha$ operator. For example, $O_2$ in \eqref{theO2} is a $D$-$(4,2)$ operator. Particularly, an operator that has no overall covariant derivative $D$ is said to be of $D$-$(0,0)$ type. We order operators strictly according to their $D$-types as
\begin{align}
&D\text{-}(i,\alpha)>D\text{-}(i',\alpha'),\ \text{if $i>i'$}\,,\\ &D\text{-}(i,\alpha)>D\text{-}(i,\alpha'),\ \text{if $\alpha<\alpha'$}\,.
\end{align}
Within each helicity sector, our operator basis enumerates all the total derivative operators from the highest to the lowest $D$-type.\footnote{Actually, one should first enumerate the $D$-type operators before classifying the operators according to helicities to make sure that all the total derivative operators are included in the basis. While it turns out that the order is irrelevant in our case.} By considering classical dimensions and Lorentz structures, it is not hard to see that an operator would not mix to any operator of lower $D$-type. Thus the $Z$-matrix and the dilatation matrix are block upper triangular according to $D$-type. There are two special operators in our basis, \emph{i.e.} $O_1$ and $O_{25}$. Each of them is the only operator of the highest $D$-type in the corresponding C-parity sector, so they cannot mix to other operators and are eigenstates of the dilatation matrix. 

Below are the operator basis. For short of notations, we drop the symbol ``tr". For example, $F_{\mu_1\mu_2}F_{\mu_1\mu_2}$ means $\text{tr}(F_{\mu_1\mu_2}F_{\mu_1\mu_2})$.

\subsection{The physical operators}

\subsubsection*{C-even}

The only length-2 operator:
\begin{flalign}
&O_1=D^6(F_{\mu_1\mu_2}F_{\mu_1\mu_2})\,.&
\end{flalign}	
Below are the length-3 operators. The $(-)^2+$ sector:
\begin{flalign}
O_2&=D^2 D_4D_5(-\eta_{45} \frac{1}{4}F_{12}F_{13}F_{23} + F_{14}F_{25}F_{12})\,,\label{theO2}&\\
O_3&=D_4D_5D_6(\eta_{56} \frac{1}{2}D_1F_{23}F_{13}F_{24} - \frac{1}{2}D_6F_{25}F_{12}F_{14} + \frac{1}{2}F_{25}D_6F_{12}F_{14}- \frac{1}{2}F_{25}F_{12}D_6F_{14})\,.
\end{flalign}
The $(-)^3$ sector:
\begin{flalign}
O_4&=\frac{D^4(F12F13F23)}{12}\,,&\\
O_5&=D_4D_5(-D_1F_{35}D_1F_{24}F_{23} -D_2F_{15}F_{34}D_1F_{23} -\frac{1}{4}F_{35}D_1F_{24}D_1F_{23} \nonumber \\
&+ \frac{3}{2}D_1F_{23}F_{12}D_4F_{35} + \frac{3}{4}F_{35}D_1F_{23}D_1F_{24})\,.
\end{flalign}	
Below are the length-4 operators. The $(-)^4$ sector:
\begin{flalign}
O_{6}&=D^2(\frac{1}{8}{F_{12}}{F_{12}}{F_{34}}{F_{34}}+\frac{1}{16}{F_{12}}{F_{34}}{F_{12}}{F_{34}}-\frac{1}{8}{F_{12}}{F_{23}}{F_{34}}{F_{14}}+\frac{3}{8}{F_{12}}{F_{34}}{F_{23}}{F_{14}})\,,&
\end{flalign}
\begin{flalign}
O_{7}&=D^2(-\frac{1}{2}{F_{12}}{F_{12}}{F_{34}}{F_{34}}-\frac{1}{4}{F_{12}}{F_{34}}{F_{12}}{F_{34}}-{F_{12}}{F_{23}}{F_{34}}{F_{14}})\,,&
\end{flalign}
\begin{flalign}
O_{8}&=D_5\bigg[-\frac{1}{2}{F_{12}}{F_{12}}{F_{34}}{D_{4}}{F_{35}}-{F_{13}}{F_{24}}{F_{35}}{D_{4}}{F_{12}}+{F_{13}}{F_{25}}{F_{34}}{D_{4}}{F_{12}}-{F_{13}}{F_{35}}{F_{24}}{D_{4}}{F_{12}}&\nonumber\\&
+{F_{24}}{F_{13}}{F_{35}}{D_{4}}{F_{12}}-{F_{24}}{F_{35}}{F_{13}}{D_{4}}{F_{12}}-\frac{1}{2}{F_{34}}{F_{12}}{F_{12}}{D_{4}}{F_{35}}+{F_{34}}{F_{12}}{F_{13}}{D_{4}}{F_{25}}\nonumber\\&
+{F_{35}}{F_{13}}{F_{24}}{D_{4}}{F_{12}}+D_6(-\frac{1}{8}{F_{12}}{F_{12}}{F_{36}}{F_{35}}+\frac{1}{4}{F_{12}}{F_{13}}{F_{36}}{F_{25}}+\frac{1}{4}{F_{12}}{F_{25}}{F_{36}}{F_{13}}\nonumber\\&
-\frac{1}{8}{F_{12}}{F_{35}}{F_{36}}{F_{12}}+\frac{1}{4}{\eta_{56}}{F_{14}}{F_{12}}{F_{23}}{F_{34}}+\frac{1}{16}{\eta_{56}}{F_{34}}{F_{12}}{F_{12}}{F_{34}}+\frac{1}{16}{\eta_{56}}{F_{34}}{F_{34}}{F_{12}}{F_{12}}\nonumber\\&
-\frac{1}{8}{F_{36}}{F_{12}}{F_{12}}{F_{35}}+\frac{1}{4}{F_{36}}{F_{13}}{F_{12}}{F_{25}}
+\frac{1}{4}{F_{36}}{F_{25}}{F_{12}}{F_{13}}-\frac{1}{8}{F_{36}}{F_{35}}{F_{12}}{F_{12}})\bigg]\,,
\end{flalign}
\begin{flalign}
O_{9}&=\frac{3}{4}F_{12}F_{12}D_{5}F_{34}D_{5}F_{34}+\frac{1}{4}F_{12}D_{5}F_{12}D_{5}F_{34}F_{34}+\frac{3}{4}F_{12}F_{34}D_{5}F_{12}D_{5}F_{34}&\nonumber\\
&+F_{12}F_{23}D_{5}F_{34}D_{5}F_{14}+2F_{12}F_{23}D_{5}F_{14}D_{5}F_{34}+2F_{12}D_{1}F_{34}D_{5}F_{23}F_{45}&\nonumber\\
&-F_{13}F_{23}D_{2}F_{45}D_{1}F_{45}\,,
\end{flalign}
\begin{flalign}
O_{10}&=-\frac{1}{2}F_{12}F_{12}D_{5}F_{34}D_{5}F_{34}-\frac{1}{2}F_{12}D_{5}F_{12}D_{5}F_{34}F_{34}-\frac{1}{2}F_{12}F_{34}D_{5}F_{12}D_{5}F_{34}&\nonumber\\
&-3F_{12}F_{23}D_{5}F_{14}D_{5}F_{34}+F_{12}D_{5}F_{23}D_{5}F_{14}F_{34}\,.
\end{flalign}
The $(-)^3+$ sector:
\begin{flalign}
O_{11}&=D_{5}D_{6}(\frac{1}{2}{F_{12}}{F_{36}}{F_{12}}{F_{35}}+{F_{13}}{F_{25}}{F_{36}}{F_{12}}-\frac{1}{4}{\eta_{56}}{F_{34}}{F_{12}}{F_{34}}{F_{12}}-{F_{36}}{F_{12}}{F_{13}}{F_{25}}&\nonumber\\
&+\frac{1}{2}{F_{36}}{F_{12}}{F_{35}}{F_{12}})\,,
\end{flalign}
\begin{flalign}
O_{12}&=D_{5}D_{6}(\frac{1}{2}{F_{12}}{F_{12}}{F_{36}}{F_{35}}-{F_{12}}{F_{13}}{F_{36}}{F_{25}}-{F_{12}}{F_{25}}{F_{36}}{F_{13}}+\frac{1}{2}{F_{12}}{F_{35}}{F_{36}}{F_{12}}&\nonumber\\&
-\frac{1}{3}{F_{12}}{F_{36}}{F_{12}}{F_{35}}-\frac{2}{3}{F_{13}}{F_{25}}{F_{36}}{F_{12}}+{F_{25}}{F_{12}}{F_{13}}{F_{36}}+{F_{25}}{F_{36}}{F_{13}}{F_{12}}\nonumber\\&
-\frac{1}{4}{\eta_{56}}{F_{34}}{F_{12}}{F_{12}}{F_{34}}+\frac{1}{6}{\eta_{56}}{F_{34}}{F_{12}}{F_{34}}{F_{12}}-\frac{1}{4}{\eta_{56}}{F_{34}}{F_{34}}{F_{12}}{F_{12}}+\frac{1}{2}{F_{36}}{F_{12}}{F_{12}}{F_{35}}\nonumber\\&
+\frac{2}{3}{F_{36}}{F_{12}}{F_{13}}{F_{25}}-\frac{1}{3}{F_{36}}{F_{12}}{F_{35}}{F_{12}}+\frac{1}{2}{F_{36}}{F_{35}}{F_{12}}{F_{12}})\,,
\end{flalign}
\begin{flalign}
O_{13}&=D_5\bigg[{D_{4}}{F_{12}}{F_{34}}{F_{13}}{F_{25}}+{D_{4}}{F_{12}}{F_{35}}{F_{13}}{F_{24}}-{D_{4}}{F_{25}}{F_{13}}{F_{34}}{F_{12}}&\nonumber\\&
+D_6(-\frac{1}{4}{F_{12}}{F_{12}}{F_{36}}{F_{35}}-\frac{1}{6}{F_{12}}{F_{36}}{F_{12}}{F_{35}}-\frac{1}{3}{F_{13}}{F_{25}}{F_{36}}{F_{12}}+\frac{1}{2}{F_{13}}{F_{36}}{F_{12}}{F_{25}}\nonumber\\&
+\frac{1}{12}{\eta_{56}}{F_{34}}{F_{12}}{F_{34}}{F_{12}}+\frac{1}{3}{F_{36}}{F_{12}}{F_{13}}{F_{25}}-\frac{1}{6}{F_{36}}{F_{12}}{F_{35}}{F_{12}}-\frac{1}{2}{F_{36}}{F_{13}}{F_{25}}{F_{12}}\nonumber\\&
+\frac{1}{4}{F_{36}}{F_{35}}{F_{12}}{F_{12}})\bigg]\,,
\end{flalign}
\begin{flalign}
O_{14}&=-\frac{1}{3}F_{12}D_{5}F_{34}F_{12}D_{5}F_{34}+\frac{2}{3}F_{13}D_{1}F_{45}F_{23}D_{2}F_{45}+\frac{2}{3}F_{13}D_{2}F_{45}F_{23}D_{1}F_{45}&\nonumber\\&
-\frac{2}{3}F_{13}D_{12}F_{45}F_{23}F_{45}\,.&
\end{flalign}	
The $(-)^2(+)^2$ sector:
\begin{flalign}
O_{15}&=D^2(\frac{1}{8}{F_{12}}{F_{34}}{F_{12}}{F_{34}}+\frac{1}{2}{F_{12}}{F_{34}}{F_{23}}{F_{14}})\,,&
\end{flalign}
\begin{flalign}
O_{16}&=D^2(-\frac{1}{4}{F_{12}}{F_{12}}{F_{34}}{F_{34}}+\frac{1}{8}{F_{12}}{F_{34}}{F_{12}}{F_{34}}-\frac{1}{2}{F_{12}}{F_{23}}{F_{34}}{F_{14}})\,,&
\end{flalign}
\begin{flalign}
O_{17}=&D_5D_6(-\frac{1}{4}{F_{12}}{F_{12}}{F_{35}}{F_{36}}-\frac{1}{2}{F_{12}}{F_{13}}{F_{36}}{F_{25}}+{F_{12}}{F_{25}}{F_{13}}{F_{36}}-\frac{1}{2}{F_{12}}{F_{25}}{F_{36}}{F_{13}}&\nonumber\\&
-\frac{1}{4}{F_{12}}{F_{35}}{F_{12}}{F_{36}}+\frac{1}{2}{F_{12}}{F_{35}}{F_{36}}{F_{12}}-\frac{1}{4}{F_{12}}{F_{36}}{F_{12}}{F_{35}}+{F_{12}}{F_{36}}{F_{13}}{F_{25}}+{F_{12}}{F_{36}}{F_{25}}{F_{13}}\nonumber\\&
-\frac{3}{4}{F_{12}}{F_{36}}{F_{35}}{F_{12}}-\frac{1}{2}{\eta_{56}}{F_{14}}{F_{12}}{F_{23}}{F_{34}}+{F_{14}}{F_{23}}{F_{12}}{F_{34}}-{F_{25}}{F_{36}}{F_{12}}{F_{13}}\nonumber\\&
+\frac{1}{4}{\eta_{56}}{F_{34}}{F_{12}}{F_{12}}{F_{34}}+\frac{1}{8}{\eta_{56}}{F_{34}}{F_{12}}{F_{34}}{F_{12}}-\frac{1}{4}{F_{35}}{F_{12}}{F_{12}}{F_{36}}+\frac{1}{4}{F_{35}}{F_{12}}{F_{36}}{F_{12}}\nonumber\\&
+\frac{1}{4}{F_{35}}{F_{36}}{F_{12}}{F_{12}}+{F_{36}}{F_{12}}{F_{13}}{F_{25}}+{F_{36}}{F_{12}}{F_{25}}{F_{13}}-\frac{3}{4}{F_{36}}{F_{12}}{F_{35}}{F_{12}}-\frac{1}{2}{F_{36}}{F_{13}}{F_{12}}{F_{25}}\nonumber\\&
+{F_{36}}{F_{13}}{F_{25}}{F_{12}}-\frac{1}{2}{F_{36}}{F_{25}}{F_{12}}{F_{13}}+{F_{36}}{F_{25}}{F_{13}}{F_{12}}-\frac{1}{2}{F_{36}}{F_{35}}{F_{12}}{F_{12}})\,,
\end{flalign}
\begin{flalign}
O_{18}&=\frac{1}{4}F_{12}F_{12}D_{5}F_{34}D_{5}F_{34}+\frac{1}{4}F_{12}D_{5}F_{12}D_{5}F_{34}F_{34}-\frac{1}{4}F_{12}F_{34}D_{5}F_{12}D_{5}F_{34}\nonumber\\&
+F_{12}F_{23}D_{5}F_{34}D_{5}F_{14}\,,\\
O_{19}&=\frac{1}{2}F_{12}F_{34}D_{5}F_{12}D_{5}F_{34}+F_{12}F_{23}D_{5}F_{14}D_{5}F_{34}+F_{12}D_{5}F_{23}D_{5}F_{14}F_{34}\,,\\
O_{20}&=-\frac{7}{4}F_{12}F_{12}D_{5}F_{34}D_{5}F_{34}+\frac{5}{4}F_{12}D_{5}F_{12}D_{5}F_{34}F_{34}-\frac{3}{4}F_{12}F_{34}D_{5}F_{12}D_{5}F_{34}&\nonumber\\&
-F_{12}F_{23}D_{5}F_{34}D_{5}F_{14}-4F_{12}F_{23}D_{5}F_{14}D_{5}F_{34}+2F_{12}D_{5}F_{23}D_{5}F_{14}F_{34}\nonumber\\&
-2F_{12}D_{1}F_{34}D_{5}F_{23}F_{45}+F_{13}F_{23}D_{1}F_{45}D_{2}F_{45}\,.
\end{flalign}
Below are the length-5 operators.
The $(-)^5$ sector:
\begin{flalign}
O_{21}&=5F_{12}F_{12}F_{34}F_{35}F_{45}+F_{12}F_{13}F_{34}F_{45}F_{25}-5F_{12}F_{13}F_{24}F_{35}F_{45}\,,&\\
O_{22}&=\frac{5}{2}F_{12}F_{12}F_{34}F_{35}F_{45}+F_{12}F_{13}F_{24}F_{45}F_{35}-3F_{12}F_{13}F_{24}F_{35}F_{45}\,.
\end{flalign}
The $(-)^3(+)^2$ sector:
\begin{flalign}
O_{23}&=\frac{3}{2}F_{12}F_{12}F_{34}F_{35}F_{45}+F_{12}F_{13}F_{34}F_{45}F_{25}-F_{12}F_{13}F_{24}F_{45}F_{35}-2F_{12}F_{13}F_{24}F_{35}F_{45}\,,&\\
O_{24}&=F_{12}F_{12}F_{34}F_{35}F_{45}-2F_{12}F_{13}F_{24}F_{45}F_{35}-2F_{12}F_{13}F_{24}F_{35}F_{45}\,.
\end{flalign}

\subsubsection*{C-odd}

The only length-3 operators of d-sector:
\begin{flalign}
O_{25}&=D_4D_5D_6(D_6F_{25}F_{12}F_{14} -F_{25}F_{12}D_6F_{14})\,.&
\label{O25}
\end{flalign}	
Below are the length-4 operators. The $(-)^4$ sector:
\begin{flalign}
O_{26}&=D_5\bigg[-{F_{12}}{F_{12}}{F_{34}}{D_{4}}{F_{35}}-2{D_{4}}{F_{12}}{F_{24}}{F_{13}}{F_{35}}+2{D_{4}}{F_{12}}{F_{25}}{F_{13}}{F_{34}}+2{D_{4}}{F_{12}}{F_{34}}{F_{13}}{F_{25}}&\nonumber\\&
+2{D_{4}}{F_{12}}{F_{34}}{F_{25}}{F_{13}}-2{D_{4}}{F_{12}}{F_{35}}{F_{13}}{F_{24}}-2{D_{4}}{F_{12}}{F_{35}}{F_{24}}{F_{13}}-2{F_{13}}{F_{24}}{F_{35}}{D_{4}}{F_{12}}\nonumber\\&
+2{F_{13}}{F_{25}}{F_{34}}{D_{4}}{F_{12}}-2{F_{13}}{F_{35}}{F_{24}}{D_{4}}{F_{12}}+2{F_{24}}{F_{13}}{F_{35}}{D_{4}}{F_{12}}-2{F_{24}}{F_{35}}{F_{13}}{D_{4}}{F_{12}}\nonumber\\&
-2{D_{4}}{F_{25}}{F_{13}}{F_{12}}{F_{34}}-2{D_{4}}{F_{25}}{F_{13}}{F_{34}}{F_{12}}-2{D_{4}}{F_{25}}{F_{34}}{F_{12}}{F_{13}}-{F_{34}}{F_{12}}{F_{12}}{D_{4}}{F_{35}}\nonumber\\&
+2{F_{34}}{F_{12}}{F_{13}}{D_{4}}{F_{25}}+2{F_{35}}{F_{13}}{F_{24}}{D_{4}}{F_{12}}+{D_{4}}{F_{35}}{F_{12}}{F_{12}}{F_{34}}+{D_{4}}{F_{35}}{F_{12}}{F_{34}}{F_{12}}\nonumber\\&
+{D_{4}}{F_{35}}{F_{34}}{F_{12}}{F_{12}}\nonumber\\&
+D_6(\frac{1}{4}{F_{12}}{F_{12}}{F_{35}}{F_{36}}-\frac{1}{2}{F_{12}}{F_{13}}{F_{25}}{F_{36}}-\frac{1}{4}{F_{12}}{F_{35}}{F_{36}}{F_{12}}+{F_{12}}{F_{36}}{F_{12}}{F_{35}}\nonumber\\&
-\frac{3}{2}{F_{12}}{F_{36}}{F_{13}}{F_{25}}-\frac{3}{2}{F_{12}}{F_{36}}{F_{25}}{F_{13}}+{F_{12}}{F_{36}}{F_{35}}{F_{12}}-\frac{1}{2}{F_{13}}{F_{36}}{F_{12}}{F_{25}}\nonumber\\&
-\frac{3}{2}{\eta_{56}}{F_{14}}{F_{23}}{F_{12}}{F_{34}}-\frac{1}{2}{F_{25}}{F_{36}}{F_{12}}{F_{13}}-\frac{1}{4}{\eta_{56}}{F_{34}}{F_{12}}{F_{12}}{F_{34}}-\frac{3}{8}{\eta_{56}}{F_{34}}{F_{12}}{F_{34}}{F_{12}}\nonumber\\&
+\frac{1}{4}{F_{35}}{F_{36}}{F_{12}}{F_{12}}+\frac{3}{4}{F_{36}}{F_{12}}{F_{12}}{F_{35}}-\frac{3}{2}{F_{36}}{F_{12}}{F_{13}}{F_{25}}-\frac{3}{2}{F_{36}}{F_{12}}{F_{25}}{F_{13}}\nonumber\\&
+{F_{36}}{F_{12}}{F_{35}}{F_{12}}-\frac{1}{2}{F_{36}}{F_{13}}{F_{25}}{F_{12}})\bigg]\,.
\end{flalign}
The $(-)^3+$ sector:
\begin{flalign}
O_{27}&=D_5D_6(-\frac{1}{2}{F_{12}}{F_{12}}{F_{36}}{F_{35}}+{F_{13}}{F_{36}}{F_{12}}{F_{25}}-{F_{36}}{F_{13}}{F_{25}}{F_{12}}+\frac{1}{2}{F_{36}}{F_{35}}{F_{12}}{F_{12}})\,,&
\end{flalign}
\begin{flalign}
O_{28}&=D_5\bigg[-2{D_{4}}{F_{12}}{F_{13}}{F_{25}}{F_{34}}-2{D_{4}}{F_{12}}{F_{34}}{F_{13}}{F_{25}}-2{D_{4}}{F_{12}}{F_{35}}{F_{13}}{F_{24}}+2{D_{4}}{F_{12}}{F_{35}}{F_{24}}{F_{13}}&\nonumber\\&
+2{D_{4}}{F_{25}}{F_{13}}{F_{34}}{F_{12}}+2{D_{4}}{F_{25}}{F_{34}}{F_{12}}{F_{13}}-{D_{4}}{F_{35}}{F_{12}}{F_{34}}{F_{12}}\nonumber\\&
+D_6(\frac{3}{4}{F_{12}}{F_{12}}{F_{36}}{F_{35}}-\frac{1}{2}{F_{12}}{F_{13}}{F_{36}}{F_{25}}-\frac{1}{2}{F_{12}}{F_{25}}{F_{36}}{F_{13}}+\frac{1}{4}{F_{12}}{F_{35}}{F_{36}}{F_{12}}\nonumber\\&
-\frac{1}{2}{F_{12}}{F_{36}}{F_{12}}{F_{35}}+{F_{12}}{F_{36}}{F_{13}}{F_{25}}+{F_{12}}{F_{36}}{F_{25}}{F_{13}}-\frac{1}{2}{F_{12}}{F_{36}}{F_{35}}{F_{12}}-{F_{13}}{F_{36}}{F_{12}}{F_{25}}\nonumber\\&
+{F_{14}}{F_{23}}{F_{12}}{F_{34}}+\frac{1}{2}{F_{25}}{F_{12}}{F_{13}}{F_{36}}+\frac{1}{2}{F_{25}}{F_{36}}{F_{13}}{F_{12}}+\frac{1}{8}{\eta_{56}}{F_{34}}{F_{12}}{F_{12}}{F_{34}}\nonumber\\&
+\frac{1}{4}{\eta_{56}}{F_{34}}{F_{12}}{F_{34}}{F_{12}}-\frac{1}{8}{\eta_{56}}{F_{34}}{F_{34}}{F_{12}}{F_{12}}-\frac{1}{4}{F_{36}}{F_{12}}{F_{12}}{F_{35}}+{F_{36}}{F_{12}}{F_{13}}{F_{25}}\nonumber\\&
+{F_{36}}{F_{12}}{F_{25}}{F_{13}}-\frac{1}{2}{F_{36}}{F_{12}}{F_{35}}{F_{12}}+{F_{36}}{F_{13}}{F_{25}}{F_{12}}-\frac{1}{4}{F_{36}}{F_{35}}{F_{12}}{F_{12}})\bigg]\,,
\end{flalign}
The $(-)^2(+)^2$ sector:
\begin{flalign}
O_{29}&=D_5\bigg[-2{D_{4}}{F_{12}}{F_{13}}{F_{35}}{F_{24}}-2{D_{4}}{F_{12}}{F_{24}}{F_{13}}{F_{35}}+2{D_{4}}{F_{12}}{F_{24}}{F_{35}}{F_{13}}+2{D_{4}}{F_{12}}{F_{34}}{F_{25}}{F_{13}}&\nonumber\\&
+2{D_{4}}{F_{12}}{F_{35}}{F_{13}}{F_{24}}-2{D_{4}}{F_{12}}{F_{35}}{F_{24}}{F_{13}}-2{D_{4}}{F_{25}}{F_{13}}{F_{34}}{F_{12}}+{D_{4}}{F_{35}}{F_{12}}{F_{34}}{F_{12}}\nonumber\\&
+D_6(-\frac{1}{4}{F_{12}}{F_{12}}{F_{36}}{F_{35}}-\frac{1}{2}{F_{12}}{F_{13}}{F_{25}}{F_{36}}+\frac{1}{2}{F_{12}}{F_{25}}{F_{13}}{F_{36}}+\frac{1}{4}{F_{12}}{F_{35}}{F_{36}}{F_{12}}\nonumber\\&
+\frac{1}{2}{F_{12}}{F_{36}}{F_{12}}{F_{35}}-\frac{1}{2}{F_{12}}{F_{36}}{F_{13}}{F_{25}}-\frac{1}{2}{F_{12}}{F_{36}}{F_{25}}{F_{13}}-\frac{1}{2}{\eta_{56}}{F_{14}}{F_{23}}{F_{12}}{F_{34}}\nonumber\\&
-{F_{25}}{F_{36}}{F_{12}}{F_{13}}-\frac{1}{8}{\eta_{56}}{F_{34}}{F_{12}}{F_{12}}{F_{34}}-\frac{1}{4}{\eta_{56}}{F_{34}}{F_{12}}{F_{34}}{F_{12}}+\frac{1}{8}{\eta_{56}}{F_{34}}{F_{34}}{F_{12}}{F_{12}}\nonumber\\&
-\frac{1}{4}{F_{35}}{F_{12}}{F_{12}}{F_{36}}+\frac{1}{4}{F_{35}}{F_{12}}{F_{36}}{F_{12}}+\frac{1}{4}{F_{35}}{F_{36}}{F_{12}}{F_{12}}+\frac{1}{2}{F_{36}}{F_{12}}{F_{12}}{F_{35}}\nonumber\\&
-\frac{1}{2}{F_{36}}{F_{12}}{F_{13}}{F_{25}}-\frac{1}{2}{F_{36}}{F_{12}}{F_{25}}{F_{13}}+\frac{1}{4}{F_{36}}{F_{12}}{F_{35}}{F_{12}}+\frac{1}{2}{F_{36}}{F_{13}}{F_{25}}{F_{12}}\nonumber\\&
+\frac{1}{2}{F_{36}}{F_{25}}{F_{13}}{F_{12}}-\frac{1}{2}{F_{36}}{F_{35}}{F_{12}}{F_{12}})\bigg]\,,
\end{flalign}
\begin{flalign}
O_{30}&=D_5\bigg[-2{D_{4}}{F_{12}}{F_{13}}{F_{24}}{F_{35}}-2{D_{4}}{F_{12}}{F_{13}}{F_{35}}{F_{24}}-2{D_{4}}{F_{12}}{F_{24}}{F_{35}}{F_{13}}+2{D_{4}}{F_{12}}{F_{25}}{F_{13}}{F_{34}}&\nonumber\\&
+2{D_{4}}{F_{12}}{F_{34}}{F_{13}}{F_{25}}+2{D_{4}}{F_{12}}{F_{34}}{F_{25}}{F_{13}}-2{D_{4}}{F_{25}}{F_{13}}{F_{12}}{F_{34}}-2{D_{4}}{F_{25}}{F_{13}}{F_{34}}{F_{12}}\nonumber\\&
-2{D_{4}}{F_{25}}{F_{34}}{F_{12}}{F_{13}}+{D_{4}}{F_{35}}{F_{12}}{F_{12}}{F_{34}}+{D_{4}}{F_{35}}{F_{12}}{F_{34}}{F_{12}}+{D_{4}}{F_{35}}{F_{34}}{F_{12}}{F_{12}}\nonumber\\&
+D_6(\frac{1}{4}{F_{12}}{F_{12}}{F_{36}}{F_{35}}+\frac{1}{2}{F_{12}}{F_{13}}{F_{25}}{F_{36}}-\frac{1}{2}{F_{12}}{F_{25}}{F_{13}}{F_{36}}-\frac{1}{4}{F_{12}}{F_{35}}{F_{36}}{F_{12}}\nonumber\\&
+\frac{1}{2}{F_{12}}{F_{36}}{F_{12}}{F_{35}}-\frac{3}{2}{F_{12}}{F_{36}}{F_{13}}{F_{25}}-\frac{3}{2}{F_{12}}{F_{36}}{F_{25}}{F_{13}}+{F_{12}}{F_{36}}{F_{35}}{F_{12}}-\frac{3}{2}{\eta_{56}}{F_{14}}{F_{23}}{F_{12}}{F_{34}}\nonumber\\&
+{F_{25}}{F_{36}}{F_{12}}{F_{13}}-\frac{3}{8}{\eta_{56}}{F_{34}}{F_{12}}{F_{12}}{F_{34}}-\frac{1}{4}{\eta_{56}}{F_{34}}{F_{12}}{F_{34}}{F_{12}}-\frac{1}{8}{\eta_{56}}{F_{34}}{F_{34}}{F_{12}}{F_{12}}\nonumber\\&
+\frac{1}{4}{F_{35}}{F_{12}}{F_{12}}{F_{36}}-\frac{1}{4}{F_{35}}{F_{12}}{F_{36}}{F_{12}}-\frac{1}{4}{F_{35}}{F_{36}}{F_{12}}{F_{12}}+\frac{1}{2}{F_{36}}{F_{12}}{F_{12}}{F_{35}}\nonumber\\&
-\frac{3}{2}{F_{36}}{F_{12}}{F_{13}}{F_{25}}-\frac{3}{2}{F_{36}}{F_{12}}{F_{25}}{F_{13}}+\frac{3}{4}{F_{36}}{F_{12}}{F_{35}}{F_{12}}-\frac{1}{2}{F_{36}}{F_{13}}{F_{25}}{F_{12}}\nonumber\\&
-\frac{1}{2}{F_{36}}{F_{25}}{F_{13}}{F_{12}}+\frac{1}{2}{F_{36}}{F_{35}}{F_{12}}{F_{12}})\bigg]\,.
\end{flalign}

\subsection{The evanescent operators}

\subsubsection*{C-even}

The 3 length-4 operators
\begin{flalign}
O_{31}&=\frac{1}{8}D_9D_{10}\bigg[(2\delta^{56789}_{3412(10)}+\delta^{34789}_{5612(10)})F_{12}F_{34}F_{56}F_{78}\bigg]\,,&\\
O_{32}&=\frac{1}{4}D_9D_{10}\bigg[(\delta^{56789}_{3412(10)}-\delta^{34789}_{5612(10)})F_{12}F_{34}F_{56}F_{78}\bigg]\,,\\
O_{33}&=\frac{1}{4}\delta^{1256(10)}_{34789}\bigg[-{D_{9}}{F_{12}}{F_{56}}{D_{10}}{F_{34}}{F_{78}}+2{D_{9}}{F_{12}}{F_{78}}{F_{56}}{D_{10}}{F_{34}}-{D_{10}}{F_{34}}{F_{78}}{D_{9}}{F_{12}}{F_{56}}\bigg]\,.
\end{flalign}
The 2 length-5 operators
\begin{flalign}
O_{34}&=\frac{1}{8}\bigg[\delta^{12347}_{569(10)8}F_{12}F_{56}F_{34}F_{9(10)}F_{78}+\delta^{12349}_{5678(10)}F_{12}F_{56}F_{78}F_{9(10)}F_{34}\bigg]\,,&\\
O_{35}&=\frac{1}{8}\bigg[-2\delta^{12347}_{569(10)8}F_{12}F_{56}F_{34}F_{9(10)}F_{78}+\delta^{12349}_{5678(10)}F_{12}F_{56}F_{78}F_{9(10)}F_{34}\bigg]\,.
\end{flalign}

\subsubsection*{C-odd}

The only length-4 operators
\begin{flalign}
O_{36}&=\frac{1}{4}D_{10}\bigg[2\delta^{2345(10)}_{67891}D_1F_{23}F_{45}F_{67}F_{89}-D_9\big(\delta^{56789}_{3412(10)}F_{12}F_{34}F_{56}F_{78}\big)\bigg]\,.&
\end{flalign}

\section{Two-loop renormalization matrix $Z^{(2)}_{\text{pp}}$}\label{zppresult}
In this appendix, we present the two-loop mixing between the physical operators. The results in the two schemes are the same, \emph{i.e.} $Z^{(2)}_{\text{pp}}=\hat{Z}^{(2)}_{\text{pp}}$. Because the $\epsilon^{-2}$ parts can be derived by the one-loop $Z$ matrix according to \eqref{z2ep2ms}, below only the $\epsilon^{-1}$ parts are presented. 

The blocks in ${Z}^{\text{even},(2)}_{\text{pp}}$ are
\begin{align}
	&{Z}^{\text{even},(2)}_{\text{pp},2\to 2}=\frac{N_c^2}{\epsilon }
	\left(
	\begin{array}{c}
		-\frac{34}{3}\\
	\end{array}
	\right)\,,
	{Z}^{\text{even},(2)}_{\text{pp},3\to 2}=\frac{N_c^2}{\epsilon}
	\left(
	\begin{array}{c}
		-\frac{1}{3} \\
		-\frac{209}{900} \\
		-1 \\
		-\frac{19}{36} \\
	\end{array}
	\right)\,,
	{Z}^{\text{even},(2)}_{\text{pp},3\to 3}=\frac{N_c^2}{\epsilon}
	\left(
	\begin{array}{cccc}
		\frac{439}{72} & 0 & \frac{3}{2} & 0 \\
		-\frac{1471}{4500} & \frac{7121}{1000} & \frac{89}{100} & 0 \\
		0 & 0 & \frac{59}{12} & 0 \\
		\frac{5923}{28800} & \frac{1531}{3200} & -\frac{655}{1152} & \frac{32459}{3456} \\
	\end{array}
	\right)\,,
\end{align}
\begin{align}
	&{Z}^{\text{even},(2)}_{\text{pp},4\to 3}=\frac{N_c^2}{\epsilon}
	\left(
	\begin{array}{cccc}
		0 & 0 & -\frac{1}{4} & 0 \\
		0 & 0 & 4 & 0 \\
		0 & 0 & \frac{1}{4} & \frac{1}{4} \\
		0 & 0 & -\frac{1}{3} & \frac{1}{9} \\
		0 & 0 & \frac{1}{6} & \frac{1}{9} \\
		\frac{29}{80} & -\frac{123}{80} & -\frac{1}{16} & -\frac{73}{144} \\
		\frac{31}{48} & -\frac{19}{16} & \frac{55}{144} & -\frac{275}{432} \\
		-\frac{1}{120} & -\frac{29}{160} & \frac{13}{288} & -\frac{29}{216} \\
		-\frac{11}{60} & \frac{3}{40} & -\frac{1}{24} & -\frac{1}{108} \\
		\frac{1}{4} & 0 & 0 & 0 \\
		-\frac{13}{18} & 0 & 0 & 0 \\
		-\frac{5}{8} & \frac{3}{8} & 0 & 0 \\
		\frac{473}{3600} & -\frac{19}{240} & 0 & 0 \\
		\frac{2}{25} & \frac{1}{40} & 0 & 0 \\
		-\frac{53}{400} & \frac{17}{80} & 0 & 0 \\
	\end{array}
	\right)\,,\\
	&{Z}^{\text{even},(2)}_{\text{pp},4\to 4}=\frac{N_c^2}{\epsilon}
	\left(
	\begin{array}{ccccccccc}
		\frac{833}{216} & \frac{449}{864} & 0 & 0 & 0 & 0 & 0 & 0 & 0 \\
		\frac{86}{27} & \frac{481}{54} & 0 & 0 & 0 & 0 & 0 & 0 & 0 \\
		\frac{907}{108} & -\frac{43}{54} & \frac{815}{72} & 0 & 0 & -\frac{391}{432} & \frac{13}{32} & \frac{31}{72} & 0 \\
		-\frac{5}{18} & \frac{43}{96} & \frac{17}{27} & \frac{3695}{216} & -\frac{10}{9} & -\frac{215}{192} & \frac{995}{1152} & -\frac{59}{32} & \frac{131}{72} \\
		\frac{143}{27} & -\frac{695}{864} & \frac{2}{27} & -\frac{373}{54} & \frac{791}{72} & \frac{1153}{1728} & -\frac{625}{1152} & \frac{509}{288} & -\frac{109}{72} \\
		\frac{551}{432} & \frac{209}{1728} & 0 & 0 & 0 & \frac{1451}{144} & -\frac{103}{64} & 0 & 0 \\
		\frac{1949}{1296} & -\frac{1525}{5184} & 0 & 0 & 0 & -\frac{7099}{1296} & \frac{8509}{1728} & 0 & 0 \\
		\frac{17}{648} & \frac{25}{5184} & \frac{17}{64} & 0 & 0 & -\frac{30197}{20736} & -\frac{5287}{6912} & \frac{12319}{1152} & 0 \\
		-\frac{221}{324} & \frac{299}{648} & -\frac{7}{48} & \frac{83}{18} & -\frac{5}{18} & \frac{2077}{5184} & \frac{173}{1728} & -\frac{127}{288} & \frac{73}{6} \\
		-\frac{1}{36} & -\frac{257}{288} & 0 & 0 & 0 & 0 & 0 & 0 & 0 \\
		-\frac{145}{54} & -\frac{329}{432} & 0 & 0 & 0 & 0 & 0 & 0 & 0 \\
		-\frac{107}{54} & \frac{43}{54} & 0 & 0 & 0 & -\frac{209}{108} & \frac{25}{18} & 0 & 0 \\
		\frac{209}{450} & \frac{499}{14400} & \frac{29}{1080} & -\frac{2299}{2160} & -\frac{121}{2700} & \frac{851}{14400} & -\frac{1909}{28800} & \frac{37}{288} & -\frac{11}{72} \\
		-\frac{1141}{2700} & -\frac{5563}{43200} & -\frac{73}{90} & \frac{1577}{720} & \frac{1531}{1800} & \frac{4543}{43200} & -\frac{779}{28800} & \frac{1}{96} & \frac{5}{24} \\
		-\frac{20371}{2700} & \frac{31997}{43200} & -\frac{10499}{720} & \frac{3661}{360} & \frac{58397}{3600} & \frac{53003}{43200} & \frac{21041}{28800} & -\frac{659}{96} & \frac{49}{12} \\
	\end{array}
	\right.\,\nonumber\\
	&\left.\qquad\qquad\qquad\qquad\qquad
	\begin{array}{cccccc}
		\frac{797}{288} & \frac{245}{96} & 0 & 0 & 0 & 0 \\
		-\frac{113}{18} & -\frac{41}{6} & 0 & 0 & 0 & 0 \\
		-\frac{497}{144} & -\frac{227}{144} & -\frac{61}{72} & 0 & 0 & 0 \\
		-\frac{1979}{3600} & \frac{1031}{2400} & -\frac{2311}{2400} & -\frac{5903}{900} & \frac{28}{45} & -\frac{331}{600} \\
		-\frac{8449}{3600} & -\frac{2867}{7200} & \frac{153}{800} & \frac{3041}{450} & \frac{1279}{360} & \frac{271}{900} \\
		-\frac{425}{192} & \frac{41}{192} & -\frac{389}{288} & 0 & 0 & 0 \\
		-\frac{251}{64} & -\frac{535}{1728} & -\frac{565}{288} & 0 & 0 & 0 \\
		-\frac{263}{192} & \frac{211}{432} & -\frac{125}{128} & 0 & 0 & 0 \\
		\frac{6319}{3240} & -\frac{2821}{2160} & \frac{12341}{12960} & -\frac{113}{15} & -\frac{367}{162} & -\frac{83}{405} \\
		\frac{953}{216} & \frac{853}{144} & 0 & 0 & 0 & 0 \\
		-\frac{2}{9} & \frac{1103}{216} & 0 & 0 & 0 & 0 \\
		\frac{11}{16} & -\frac{533}{144} & \frac{299}{54} & 0 & 0 & 0 \\
		-\frac{1583}{13500} & \frac{6623}{6000} & -\frac{23039}{108000} & \frac{701849}{54000} & \frac{92}{675} & -\frac{3899}{27000} \\
		-\frac{254219}{108000} & \frac{54473}{36000} & -\frac{23063}{216000} & -\frac{353}{125} & \frac{273949}{21600} & -\frac{53341}{108000} \\
		\frac{31951}{3000} & \frac{176}{125} & \frac{319699}{72000} & \frac{63643}{3000} & -\frac{125777}{7200} & \frac{854629}{108000} \\
	\end{array}
	\right)\,,\\
	&{Z}^{\text{even},(2)}_{\text{pp},5\to 4}=\frac{N_c^2}{\epsilon}
	\left(
	\begin{array}{ccccccccccccccc}
		0 & 0 & \frac{5}{2} & -\frac{5}{2} & -\frac{5}{2} & 0 & 0 & 0 & 0 & 0 & 0 & 0 & 0 & 0 & 0 \\
		0 & 0 & \frac{7}{4} & -\frac{5}{4} & -\frac{7}{4} & 0 & 0 & 0 & 0 & 0 & 0 & 0 & 0 & 0 & 0 \\
		0 & 0 & 0 & 0 & 0 & \frac{1}{96} & 0 & \frac{1}{16} & -\frac{1}{32} & \frac{3}{16} & 0 & \frac{3}{32} & -\frac{1}{8} & -\frac{17}{16} & -\frac{1}{16} \\
		0 & 0 & 0 & 0 & 0 & -\frac{1}{144} & \frac{1}{6} & -\frac{17}{24} & \frac{17}{48} & -\frac{65}{216} & 0 & -\frac{65}{432} & \frac{3}{4} & \frac{91}{216} & -\frac{13}{216} \\
	\end{array}
	\right)\,,\\
	&{Z}^{\text{even},(2)}_{\text{pp},5\to 5}=\frac{N_c^2}{\epsilon}
	\left(
	\begin{array}{cccc}
		-\frac{409}{36} & \frac{1045}{36} & \frac{1535}{144} & \frac{1135}{288} \\
		-\frac{1123}{72} & \frac{749}{24} & \frac{107}{16} & \frac{755}{288} \\
		-\frac{85}{72} & \frac{121}{72} & \frac{13771}{1728} & \frac{2657}{1152} \\
		-\frac{11}{18} & \frac{3}{4} & \frac{77}{32} & \frac{14359}{1728} \\
	\end{array}
	\right)\,.
\end{align}
The blocks in ${Z}^{\text{odd},(2)}_{\text{pp}}$ are
\begin{align}
	&{Z}^{\text{odd},(2)}_{\text{pp},3\to 3}=\frac{N_c^2}{\epsilon}
	\left(
	\begin{array}{c}
		\frac{103}{18}\\
	\end{array}
	\right)\,,
	{Z}^{\text{odd},(2)}_{\text{pp},4\to 3}=\frac{N_c^2}{\epsilon}
	\left(
	\begin{array}{c}
		0 \\
		-\frac{3}{32} \\
		-\frac{9}{32} \\
		0 \\
		\frac{1}{6} \\
	\end{array}
	\right)\,,
	{Z}^{\text{odd},(2)}_{\text{pp},4\to 4}=\frac{N_c^2}{\epsilon}
	\left(
	\begin{array}{ccccc}
		\frac{289}{24} & \frac{17}{24} & -\frac{73}{288} & -\frac{77}{360} & -\frac{11}{45} \\
		0 & \frac{3125}{384} & 0 & 0 & 0 \\
		-\frac{1}{24} & -\frac{353}{288} & \frac{13267}{1152} & -\frac{37}{144} & \frac{7}{9} \\
		\frac{2797}{720} & -\frac{1}{8} & \frac{541}{1440} & \frac{14867}{2160} & -\frac{19}{60} \\
		-\frac{413}{720} & \frac{5}{6} & \frac{89}{160} & -\frac{13}{24} & \frac{10729}{1080} \\
	\end{array}
	\right)\,.
\end{align}

\end{appendix}

\providecommand{\href}[2]{#2}\begingroup\raggedright\endgroup

\end{document}